\documentclass[
reprint,
superscriptaddress,
amsmath,
amssymb,
prx,
aps,
floatfix,
]{revtex4-2}

\usepackage[T1]{fontenc}

\usepackage{mathtools}
\usepackage{amsthm}

\usepackage{bm}
\usepackage{bbm}
\usepackage{stmaryrd}

\usepackage{graphicx}
\usepackage[table]{xcolor}
\usepackage{dcolumn}
\usepackage{rotating}

\usepackage{array}
\usepackage{booktabs}
\usepackage{tabularx}
\usepackage{longtable}
\usepackage{makecell}
\usepackage{hhline}

\newcolumntype{C}{>{\centering\arraybackslash}X}

\usepackage{tikz}
\usepackage{pgfplots}
\usepackage{tkz-graph}
\usepackage{quantikz}

\pgfplotsset{
  compat=newest,
  plot coordinates/math parser=false
}

\usetikzlibrary{
  arrows.meta,
  calc,
  decorations.markings,
  matrix,
  plotmarks,
  positioning
}

\usepgfplotslibrary{patchplots}

\usepackage{ifthen}
\usepackage{xifthen}
\usepackage[normalem]{ulem}

\usepackage{hyperref}
\hypersetup{hidelinks}

\newtheorem{theorem}{Theorem}

\newtheorem{remark}[theorem]{Remark}

\newtheorem*{example*}{Example}

\newif\ifnotes
\notestrue

\newcommand{\llbr}{[\![}
\newcommand{\rrbr}{]\!]}

\allowdisplaybreaks


\begin{document}


\title{Towards Block-Level Fault-Tolerant Quantum Simulation\\ on Small High-Rate Non-CSS Codes}


\author{Zhuangzhuang Chen}
\email{zhuangzhuangchen@arizona.edu}

\author{Narayanan Rengaswamy}%
\email{narayananr@arizona.edu}

\affiliation{%
School of Electrical, Computing and Software Engineering,\\
University of Arizona, Tucson, AZ 85721, USA
}

\date{\today}

\begin{abstract}
Small high-rate non-CSS stabilizer codes provide compact platforms for
encoded quantum computation, but their mixed-Pauli checks and lack of
a sufficiently rich known set of native transversal logical gates
complicate the implementation of fault-tolerant dynamics. Structured
block-level constructions offer a complementary approach by mapping an
entire logical block to a physical circuit rather than compiling
it gate by gate into separately protected logical primitives. In this
work, we investigate the opportunities and limitations of this
approach using the high-rate \(\llbr 8,3,3\rrbr\) non-CSS code and
logical Trotter circuits as a representative testbed. We first
construct flagged syndrome-extraction circuits and establish a
circuit-level memory pseudo-threshold near
\(1.5\times10^{-3}\). We then apply our
symplectic-transvection construction, which maps a logical Trotter
circuit to a physical circuit with the same block pattern for any
stabilizer code. Although this mapping preserves the intended unitary
algebraically, the encoded Trotter circuits exhibit a pronounced
asymmetry between the logical-\(X\) and logical-\(Z\) failure
channels. Single-fault analysis identifies the underlying mechanism:
a fault on the shared parity ancilla can propagate through the
uncomputation network into an undetectable logical operator, reducing
the effective circuit distance in the affected sector. We evaluate
flag-conditioned recovery, biased-noise decoding, CliNR resource
verification, flag postselection, and asymmetric gate-noise models.
These methods suppress substantial classes of propagated faults, but
the realistic configurations studied do not simultaneously suppress
both logical sectors. A diagnostic protected limit that removes the
identified malignant first-order locations restores pseudo-threshold
behavior in both sectors, with performance approaching that of the
memory experiment. These results demonstrate both the potential of
block-level logical constructions for non-CSS codes without rich
native transversal gate sets and the additional joint protection of
the parity network, analog rotation, and recovery required to
preserve fault-tolerant distance.
\end{abstract}

\maketitle













\section{Introduction}
\label{sec:Introduction}

Quantum hardware and quantum-error-correction experiments are
progressing from demonstrations of isolated ingredients toward repeated
syndrome extraction, logical operations, and small algorithmic
workloads~\cite{google2023suppressing,google2025below,
bluvstein2024logical,froland2026realizing}. Most experimental and
architectural studies have nevertheless focused on error-detecting
codes, small CSS codes, or surface-code-derived
constructions~\cite{calderbank1996good,steane1996error,
steane1997active,horsman2012surface,litinski2019game}. This emphasis is
well motivated by the separated $X$- and $Z$-check structure and the
availability of mature logical primitives, but it leaves the
circuit-level behavior of small non-CSS codes comparatively unexplored.
Such codes are attractive near-term testbeds because their favorable
finite-size rates can make nontrivial logical workloads accessible with
limited hardware, while revealing whether a protocol relies on
CSS-specific measurement structure, transversal gates, or permutation
automorphisms~\cite{eastin2009restrictions,grassl2013leveraging}.

The $\llbr 8,3,3\rrbr$ stabilizer code considered here is a compact,
high-rate example. It encodes three logical qubits into eight physical
qubits, giving $k/n=3/8$, while retaining distance three and therefore
correcting an arbitrary single-qubit error. Its stabilizer generators
contain mixed $X$, $Y$, and $Z$ components. Gottesman's construction
provides the code description, and Chao and Reichardt used the code in
developing flag-based fault-tolerant error
correction~\cite{gottesman1997stabilizer,chao2018quantum}. However, its
flagged extraction circuits, flag-conditioned recovery tables, and
circuit-level performance have not previously been presented together.
Moreover, the code is not known to possess a sufficiently rich native
set of transversal, fold-transversal, or automorphism-based logical
gates for quantum simulation. It therefore separates two questions
that are easily conflated: whether an encoded block can be protected as
a memory, and whether useful logical dynamics can be executed without
reducing its effective circuit distance.

Quantum simulation provides a natural setting in which to study this
distinction. Accurately representing and evolving correlated many-body
states is a central challenge in physics and chemistry, with
applications to molecular electronic structure, reaction mechanisms,
catalysis, and pharmaceutical research
~\cite{georgescu2014quantum,mcardle2020quantum,
reiher2017elucidating,blunt2022perspective,bauer2020quantum}. Hamiltonian simulation is
therefore a recurring primitive in quantum algorithms for dynamics and
energy estimation. Product formulas provide a particularly direct
realization: after decomposing a Hamiltonian into Pauli terms, they
approximate its time evolution by repeated sequences of Pauli
rotations, producing Trotter blocks whose depth grows with simulated
time and target accuracy~\cite{lloyd1996universal,childs2021theory}.
This transparent circuit structure makes Trotterization attractive for
algorithm-tailored implementation, while its repeated multiqubit
interactions make fault propagation a central concern. Quantum error
correction offers a principled route to reliable long-depth
computation~\cite{aliferis2005quantum,terhal2015quantum}, but a complete
implementation must protect syndrome extraction, logical Clifford
operations, non-Clifford resources, decoding, and state conversion.
Because no stabilizer code admits a universal transversal gate
set~\cite{eastin2009restrictions}, general-purpose schemes employ
resources such as magic-state injection and
distillation~\cite{bravyi2005universal,litinski2019magic}, code switching
or gauge fixing~\cite{paetznick2013universal,jochym2014using,
bombin2015gauge}, and measurement-based lattice
surgery~\cite{horsman2012surface,litinski2019game}. Generalized surgery
similarly enables logical measurements on QLDPC codes using long-range
connectivity and bridging ancillas~\cite{cohen2022low,cross2024improved}.
These methods are universal and highly developed, but their ancillary
systems, repeated measurements, routing, and conversion steps motivate
a complementary question: can the repeated structure of a simulation
kernel be protected and implemented directly, rather than decomposed
gate by gate into a universal library of logical primitives?

Partially fault-tolerant architectures address a related question by
allocating different levels of protection to operations with unequal
costs and fault-propagation risks. STAR combines error-corrected
Clifford operations with space--time-efficient analog
rotations~\cite{akahoshi2024partially,akahoshi2025compilation}, while
Flexion dynamically converts between bare qubits and encoded patches in
response to trapped-ion hardware asymmetry~\cite{yin2025flexion}.
CliNR instead verifies resource states for gate-teleported Clifford
circuits~\cite{delfosse2025low}. Recent Ising-model experiments on IBM
hardware combined fault-tolerant syndrome extraction with
non-fault-tolerant logical operations in the $\llbr4,2,2\rrbr$ Iceberg
code and obtained observable-dependent improvements over unencoded
circuits~\cite{froland2026realizing}. Algorithmic fault tolerance
further shows that syndrome information and correctness can sometimes
be treated across an algorithm rather than by closing every gadget
independently~\cite{zhou2024algorithmic}. Pieceable fault tolerance
offers another route for selected nontransversal gates by interleaving
restricted interactions with intermediate error
correction~\cite{yoder2016universal,yoder2018practical}; its relation to
the parity-mediated Trotter circuits considered here is discussed in
Sec.~\ref{sec:discussion}.

Our approach changes the implementation granularity at which the
simulation circuit is synthesized. In earlier work, we introduced a
\emph{solve-and-stitch} construction that exploited complete Clifford
Trotter blocks on the $\llbr n,n-2,2\rrbr$ Iceberg-code family and used
low-overhead flag gadgets to protect them
fault-tolerantly~\cite{chen2025tailoring,chen2024tailoring}. We
subsequently showed that symplectic transvections provide a
code-independent logical-to-physical mapping for Trotter circuits on
arbitrary stabilizer codes~\cite{chen2025fault}. If an unencoded Pauli
operator $E$ is mapped by the encoding isometry to a physical
representative $\overline{E}$, then
\begin{equation}
    R_E(\theta)
    \longmapsto
    R_{\overline{E}}(\theta).
    \label{eq:intro_trotter_pattern_mapping}
\end{equation}
Thus, the logical and physical circuits share the same Trotter pattern:
each unencoded Pauli support is replaced by its encoded representative,
which can be applied to any stabilizer code. The construction is
\emph{block-level} because a complete logical Trotter block is mapped
to a unitary physical block within one encoding, rather than compiled
gate by gate into separately protected logical operations. 
Our approach was recently applied by a team from qBraid, IonQ, and NVIDIA
to demonstrate beyond break-even performance of encoded fermionic quantum
simulation on trapped-ion hardware~\cite{brown2025efficient,brown2026mid}.
This algebraic mapping does not, by itself, guarantee circuit-level fault
tolerance: a physical representative may have high weight, and a
single fault in its parity network may be closed by the remaining
circuit into an undetectable logical operator.

We use the $\llbr8,3,3\rrbr$ code as a compact and deliberately
demanding circuit-level testbed for this general framework. Its
mixed-Pauli checks remove CSS-specific simplifications, its high rate
supports three-logical-qubit dynamics within a small block, and its
size permits explicit single-fault analysis. The underlying
logical-to-physical mapping is not specialized to this code, and our goal
is not to infer large-code scaling from one example. Rather, we ask
when fault-tolerant memory protection remains compatible with the
shared logical--physical Trotter structure, identify the mechanisms
that destroy this compatibility, and determine what additional
protection a practical realization would require.

The main contributions are as follows:

\begin{enumerate}
    \item We establish code-capacity, phenomenological-noise, and
    circuit-level memory baselines for the $\llbr8,3,3\rrbr$ code. We
    construct five two-ancilla flagged syndrome-extraction circuits and
    their recovery tables following the Chao--Reichardt
    framework~\cite{chao2018quantum}. Under the final ideal-cleanup
    boundary convention used throughout the circuit study, the memory
    experiment exhibits a pseudo-threshold near
    $1.5\times10^{-3}$; see Secs.~\ref{sec:833_memory}
    and~\ref{sec:833_flagged_qec}.

    \item We apply the code-independent Trotter mapping and evaluate
    propagated logical observables. Single-fault analysis identifies a
    circuit-distance-one mechanism in which a phase-containing fault on
    the parity ancilla is converted by the closing CNOT ladder into a
    full logical-$Z$ operator with trivial syndrome. This mechanism
    contributes to the strong logical-sector asymmetry of the direct
    encoded circuit; see Sec.~\ref{sec:833_flagged_qec}.

    \item We compare flag-conditioned recovery, $Z$-biased Trotter
    noise, CliNR resource verification, and direct flag postselection.
    Several configurations strongly suppress one logical sector, but
    none of the tested noisy circuits suppresses both sectors
    simultaneously under the stated conventions. Fault enumeration
    shows that distinct propagated errors can share the same flag and
    syndrome record, so the asymmetry is not solely a decoder artifact;
    see Secs.~\ref{sec:833_clinr}
    and~\ref{sec:833_flag_postselection}.

    \item We separate the error scales of the Clifford parity network
    and central phase rotation. At fixed
    $p_{\mathrm{CNOT}}=10^{-5}$ (e.g., achievable through gate teleportation with highly purified Bell pairs), the direct encoded circuit exhibits a
    low-noise CNOT floor, while its logical-$Z$ failure rate approaches
    $2p_{\mathrm{phase}}/3$ at larger phase-gate error. Adding a noisy
    Chao--Reichardt flagged syndrome extraction round exposes the cost of dynamic recovery. In a
    diagnostic protected limit, ideal phase and flag-gadget couplings
    remove the identified first-order Trotter mechanisms and restore a
    pseudo-threshold in both sectors, leaving performance close to the
    memory benchmark. This is not a hardware-ready protocol; it defines
    the protection that a practical construction must reproduce. See
    Sec.~\ref{sec:833_asymmetric_noise}.
\end{enumerate}

Together, these results show that preserving a logical Trotter pattern
under encoding does not necessarily preserve fault-tolerant distance.
The relevant circuit-level criterion is whether available syndrome and
flag information remains sufficient before a propagated fault is
closed into a logical operator. The $\llbr8,3,3\rrbr$ study therefore
provides both a concrete non-CSS benchmark and a design criterion for
block-level fault tolerance on more general stabilizer codes. We
discuss implications and open problems in Sec.~\ref{sec:discussion}
and summarize the conclusions in Sec.~\ref{sec:conclusion}.

\section{Preliminary background}
\label{sec:preliminaries}

This section fixes the notation used for stabilizer codes, Pauli
rotations, and circuit-level fault tolerance. We include only the
definitions needed for the subsequent \(\llbr 8,3,3\rrbr\) study.

\subsection{Pauli operators and stabilizer codes}

Let \(\mathcal{P}_n\) denote the \(n\)-qubit Pauli group. Up to the
phases \(\{\pm1,\pm i\}\), every element can be written as
\begin{equation}
    P
    =
    \bigotimes_{j=1}^{n} X_j^{x_j} Z_j^{z_j},
    \qquad
    (\boldsymbol{x}\mid\boldsymbol{z})
    \in \mathbb{F}_2^{2n}.
    \label{eq:pauli_binary_representation}
\end{equation}
Two Pauli operators with binary representations
\((\boldsymbol{x}\mid\boldsymbol{z})\) and
\((\boldsymbol{x}'\mid\boldsymbol{z}')\) commute if and only if their
binary symplectic inner product vanishes,
\begin{equation}
    \boldsymbol{x}\!\cdot\!\boldsymbol{z}'
    +
    \boldsymbol{z}\!\cdot\!\boldsymbol{x}'
    =0
    \pmod 2.
    \label{eq:binary_symplectic_commutation}
\end{equation}
This representation provides the algebraic connection between Pauli
propagation and the symplectic-transvection construction used in our
logical-to-physical Trotter mapping~\cite{gottesman1997stabilizer,chen2025fault}.

An \(\llbr n,k,d\rrbr\) stabilizer code is specified by an Abelian
subgroup \(\mathcal{S}\subset\mathcal{P}_n\) generated by
\(n-k\) independent commuting Pauli operators and not containing
\(-I\). The codespace is the simultaneous \(+1\) eigenspace of
\(\mathcal{S}\) and therefore has dimension \(2^k\). For an ordered
generating set \((g_1,\ldots,g_{n-k})\), a Pauli error \(E\) produces
a binary syndrome whose \(i\)-th component records whether \(E\)
commutes or anticommutes with \(g_i\). A recovery rule uses this
syndrome, and any additional measurement record such as flag
outcomes, to choose a Pauli correction.

Let \(\mathcal{N}(\mathcal{S})\) denote the Pauli normalizer of the
stabilizer group. Elements of
\(\mathcal{N}(\mathcal{S})\setminus\mathcal{S}\) preserve the
codespace but act nontrivially on the encoded information; equivalence
classes in \(\mathcal{N}(\mathcal{S})/\mathcal{S}\) therefore define
the logical Pauli operators. The code distance is
\begin{equation}
    d
    =
    \min_{P\in
    \mathcal{N}(\mathcal{S})\setminus\mathcal{S}}
    \operatorname{wt}(P).
    \label{eq:stabilizer_code_distance}
\end{equation}
Thus, every Pauli error of weight at most
\(\lfloor(d-1)/2\rfloor\) is correctable. Importantly, this static
code distance does not by itself determine the fault tolerance of a
circuit implementing an encoded operation: a single circuit fault
may propagate into a higher-weight error or directly into a logical
coset~\cite{gottesman1997stabilizer,terhal2015quantum}.

\subsection{Hamiltonian simulation and Pauli-rotation circuits}

For digital Hamiltonian simulation, the target Hamiltonian is
expanded in the Pauli basis,
\begin{equation}
    H
    =
    \sum_{j=1}^{m} h_j P_j,
    \qquad
    P_j\in\mathcal{P}_n,
    \label{eq:hamiltonian_pauli_decomposition}
\end{equation}
and the desired time evolution is \(U(t)=e^{-itH}\). When the terms
do not mutually commute, a product formula approximates this
evolution by a sequence of Pauli rotations. For example, the
first-order formula with \(r\) Trotter steps is
\begin{equation}
    U(t)
    \approx
    \left[
        \prod_{j=1}^{m}
        \exp\!\left(-i\frac{t h_j}{r}P_j\right)
    \right]^{r}.
    \label{eq:first_order_product_formula}
\end{equation}
The approximation error is governed by the noncommutativity of the
Hamiltonian terms and decreases as the number of product-formula
steps is increased~\cite{lloyd1996universal,childs2021theory}.

\begin{figure}[t]
    \centering
    \includegraphics[width=0.75\columnwidth]
        {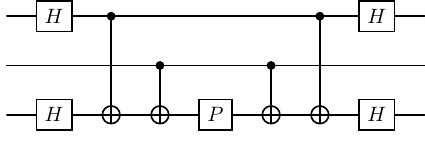}
    \caption{A three-qubit compute--rotate--uncompute circuit for the
    Clifford Pauli rotation
    \(\exp(-i\frac{\pi}{4}X_1Z_2X_3)\), up to a global phase. The Hadamard
    gates map the \(X\) factors to the \(Z\) basis, and the CNOT ladder
    computes the corresponding parity onto the third qubit. The
    central phase gate \(P=\operatorname{diag}(1,i)\) applies the
    rotation, after which the parity computation and basis changes are
    reversed.}
    \label{fig:three_qubit_trotter}
\end{figure}

We write a Pauli rotation as
\begin{equation}
    R_P(\theta)
    =
    \exp\!\left(-i\frac{\theta}{2}P\right)
    =
    \cos\!\left(\frac{\theta}{2}\right)I
    -i\sin\!\left(\frac{\theta}{2}\right)P.
    \label{eq:pauli_rotation_definition}
\end{equation}
For a multiqubit Pauli string, a standard implementation first maps
its local \(X\) and \(Y\) factors to the \(Z\) basis, computes the
corresponding parity onto a designated target or ancilla using an
entangling-gate ladder, applies a single-qubit \(Z\) rotation, and
then reverses the parity computation and basis changes. As a concrete
three-qubit example, Fig.~\ref{fig:three_qubit_trotter} implements the
Clifford Pauli rotation
\begin{equation}
    \exp\left(-i\frac{\pi}{4}X_1Z_2X_3\right)
\end{equation}
up to a global phase. Hadamard gates map the \(X\) factors to the
\(Z\) basis, the CNOT ladder computes the resulting three-qubit parity
onto the third qubit, and the central phase gate applies the required
rotation before the circuit is uncomputed. This
compute--rotate--uncompute pattern is the Trotter primitive studied
throughout this work. Although it is algebraically exact for the
individual Pauli rotation, faults on the parity target can propagate
through the uncomputation ladder and become correlated data errors.

Let \(\mathcal{E}\) denote an encoding circuit, and let \(Q\) act on
the logical input qubits before encoding. A physical representative
of the corresponding logical Pauli is
\begin{equation}
    \overline{Q}
    =
    \mathcal{E}
    \bigl(Q\otimes I\bigr)
    \mathcal{E}^{\dagger},
    \label{eq:encoded_pauli_representative}
\end{equation}
where representatives that differ by a stabilizer have the same
action on the codespace. Conjugation commutes with the operator
exponential, so
\begin{equation}
    \mathcal{E}
    \bigl(R_Q(\theta)\otimes I\bigr)
    \mathcal{E}^{\dagger}
    =
    R_{\overline{Q}}(\theta).
    \label{eq:encoded_pauli_rotation}
\end{equation}
Consequently, the logical and physical circuits share the same
sequence of Pauli rotations after each logical Pauli is replaced by
a physical representative~\cite{chen2025fault}. This is the
code-independent structural mapping exploited here. It does not,
however, guarantee that the parity circuit realizing
\(R_{\overline{Q}}(\theta)\) is fault tolerant.

\subsection{Circuit-level fault tolerance and pseudo-thresholds}

For a distance-three code, an encoded gadget is fault tolerant at
first order only if every single physical fault either leaves a
correctable data error or produces a measurement record that
distinguishes the appropriate recovery class. If one physical fault
can produce a nontrivial logical operator with a record identical to
that of the identity, the gadget has effective circuit-level distance
one in that logical sector, regardless of the static code distance.
Flag qubits are intended to enlarge the available measurement record
by detecting ancilla faults capable of producing correlated data
errors~\cite{chao2018quantum}.

Because the present work studies a fixed small code rather than a
distance-scaling family, performance is reported using
\emph{pseudo-thresholds}. For a specified circuit, noise model,
decoder, and reference process, a pseudo-threshold is a crossing
point error rate \(p^*\) satisfying
\begin{equation}
    P_L(p^*)
    =
    P_{\mathrm{ref}}(p^*), \ \text{and}\ 
    P_L(p)
    <
    P_{\mathrm{ref}}(p) \ \text{for}\ p < p^*,
    \label{eq:pseudothreshold_definition}
\end{equation}
where \(P_L\) is the encoded logical failure probability. Depending
on the experiment, \(P_{\mathrm{ref}}\) is either the corresponding
unencoded-circuit failure probability or the explicitly stated line
\(P_L=p\). A pseudo-threshold is therefore a finite-size benchmark
and should not be confused with an asymptotic accuracy threshold for
a family of codes with increasing distance.

\section{Circuit-Level Study of the
$\llbr 8,3,3\rrbr$ Code}
\label{sec:833}

\subsection{Code description and memory benchmarks}
\label{sec:833_memory}

We begin our circuit-level study with the
$\llbr 8,3,3\rrbr$ stabilizer code, which encodes $k=3$ logical
qubits into $n=8$ physical qubits and has distance $d=3$. The code
is non-CSS: several stabilizer generators contain mixed Pauli
operators, and the logical Pauli operators are not separated into
purely $X$-type and $Z$-type sectors. At the same time, its encoding
rate is $k/n=3/8$, which is relatively high for a small
distance-three code. These properties make the code a useful
small-scale setting in which to study circuit-level fault tolerance
for general stabilizer codes.

A generating set for the stabilizer group is
\begin{equation}
    \mathcal{S}_{833}
    =
    \left\langle
        S_1,S_2,S_3,S_4,S_5
    \right\rangle ,
    \label{eq:833_stabilizer_group}
\end{equation}
where the stabilizer generators and one choice of logical Pauli
representatives are listed in Table~\ref{tab:833_code}. Since there
are five independent stabilizer generators acting on eight physical
qubits, the codespace has dimension $2^{8-5}=2^3$.

\begin{table}[h]
    \centering
    \caption{
        Stabilizer generators and logical Pauli representatives for
        the $\llbr 8,3,3\rrbr$ code.
    }
    \label{tab:833_code}
    \begin{tabular}{cl}
        \toprule
        Operator & Pauli representation \\
        \midrule

        $S_1$
        &
        $X_1X_2X_3X_4X_5X_6X_7X_8$
        \\

        $S_2$
        &
        $Z_1Z_2Z_3Z_4Z_5Z_6Z_7Z_8$
        \\

        $S_3$
        &
        $Z_3Y_4X_5Z_6Y_7X_8$
        \\

        $S_4$
        &
        $Z_2X_3X_5Y_6Z_7Y_8$
        \\

        $S_5$
        &
        $X_2Z_4Z_5X_6Y_7Y_8$
        \\

        \midrule

        $\overline{X_1}$
        &
        $X_4X_5X_7X_8$
        \\

        $\overline{X_2}$
        &
        $X_3Z_4Z_5X_6$
        \\

        $\overline{X_3}$
        &
        $Z_1Z_2X_6X_7$
        \\

        $\overline{Z_1}$
        &
        $Z_2X_3Z_5X_8$
        \\

        $\overline{Z_2}$
        &
        $Z_1Z_5Z_6Z_7$
        \\

        $\overline{Z_3}$
        &
        $Z_1Z_2Z_4Z_7$
        \\

        \bottomrule
    \end{tabular}
\end{table}

The code is non-degenerate, so the 24 single-qubit Pauli errors
\begin{equation}
    \mathcal{E}_1
    =
    \left\{
        X_i,Y_i,Z_i
        \,\middle|\,
        i=1,\ldots,8
    \right\}
    \label{eq:833_weight_one_errors}
\end{equation}
produce distinct nonzero syndromes for any fixed independent
generating set. We therefore use a weight-one lookup-table decoder,
\begin{equation}
    \mathcal{D}_{\mathrm{LUT}}
    \bigl(
        \boldsymbol{s}(E)
    \bigr)
    =
    E,
    \qquad
    E\in\mathcal{E}_1,
    \label{eq:833_weight_one_decoder}
\end{equation}
with the trivial syndrome mapped to the identity. Syndromes that do
not correspond to weight-one Pauli errors are declared outside the
weight-one lookup table. No recovery operation is assigned to such
syndromes, and the corresponding trials are counted as decoding
failures. This treatment explains the close agreement with the
analytical reference curve, which counts all events containing two or
more data-qubit errors as failures.

For consistency with the subsequent circuit-level simulations, all
numerical experiments use the syndrome convention associated with
the five extraction generators introduced in
Sec.~\ref{sec:833_flagged_qec}. The explicit weight-one syndrome
assignments are listed in Table~\ref{tab:833_weight1_lut}. This choice
only fixes the binary representation of the syndrome and does not
change the stabilizer group, codespace, or set of correctable errors.

Before introducing noisy syndrome-extraction circuits and logical
Trotter evolution, we benchmark the $\llbr 8,3,3\rrbr$ code as a
quantum memory. These experiments verify the state preparation,
syndrome calculation, weight-one recovery, and logical-error
classification used in the later circuit-level simulations. They
also establish a memory baseline against which the additional faults
introduced by noisy syndrome extraction and logical evolution can be
compared.

We first consider the code-capacity noise model. All code-capacity and phenomenological-noise data points reported in
this subsection are estimated from \(N=20,000\) Monte Carlo shots
per sampled parameter setting. Each data qubit
independently undergoes the depolarizing channel
\begin{equation}
    \mathcal{N}_{p}(\rho)
    =
    (1-p)\rho
    +
    \frac{p}{3}
    \left(
        X\rho X
        +
        Y\rho Y
        +
        Z\rho Z
    \right),
    \label{eq:833_depolarizing_channel}
\end{equation}
while state preparation, syndrome extraction, and recovery are
assumed to be noiseless. Thus, each nonidentity single-qubit Pauli
error occurs with probability $p/3$.

Because the code corrects every weight-one Pauli error, failure of the
weight-one decoder requires errors on at least two data qubits. The
probability that two or more of the eight data qubits are affected is
\begin{align}
    P_{\geq 2}(p)
    &=
    1-(1-p)^8-8p(1-p)^7
    \nonumber\\
    &=
    28p^2+O(p^3).
    \label{eq:833_two_or_more_errors}
\end{align}
Not every error of weight two or larger necessarily produces a
logical failure, but Eq.~\eqref{eq:833_two_or_more_errors} provides a
useful reference for the onset of errors beyond the guaranteed
correction capability of the code.

\begin{figure}[h]
    \centering
    \includegraphics[
        width=\columnwidth
    ]{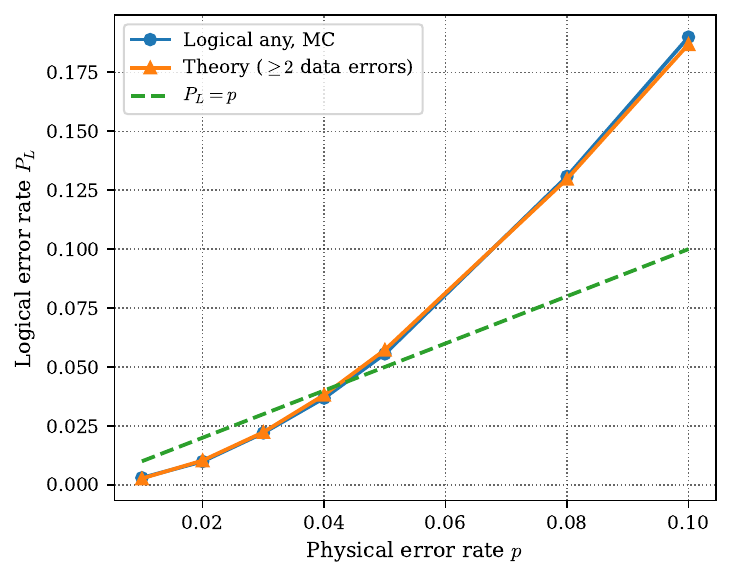}
    \caption{
        Logical error rate of the $\llbr 8,3,3\rrbr$ code under the
        code-capacity depolarizing-noise model. The Monte Carlo result
        is compared with the probability of two or more data-qubit
        errors from Eq.~\eqref{eq:833_two_or_more_errors} and with the
        unencoded reference $P_L=p$. The encoded and unencoded curves
        cross near $p\simeq0.04$, corresponding to a code-capacity
        pseudo-threshold of approximately $4\%$. Each point is estimated from \(N=20,000\) Monte Carlo shots.
    }
    \label{fig:833_code_capacity}
\end{figure}

Figure~\ref{fig:833_code_capacity} compares the Monte Carlo logical
error rate with the probability in
Eq.~\eqref{eq:833_two_or_more_errors}. Their close agreement confirms
the suppression of all weight-one errors and provides a consistency
check of the code implementation and decoder. At small $p$, the
logical error rate exhibits the expected quadratic scaling. The
encoded curve crosses the unencoded reference $P_L=p$ near
$p\simeq0.04$, giving a code-capacity pseudo-threshold of
approximately $4\%$. We use the term \emph{pseudo-threshold} because
the comparison involves a single finite-size code rather than a
family of codes with increasing distance.

We next consider a phenomenological measurement-noise model. At the
beginning of each memory experiment, the data qubits undergo the
channel in Eq.~\eqref{eq:833_depolarizing_channel} with error
probability $p_{\mathrm{data}}$. This data-noise layer is applied only
once, so the underlying data-error pattern remains fixed during the
repeated syndrome measurements.

Let $\boldsymbol{s}=(s_1,\ldots,s_5)$ denote the ideal syndrome in the
extraction-generator ordering defined in
Sec.~\ref{sec:833_flagged_qec}. During measurement round $r$, the
reported value of syndrome bit $j$ is
\begin{equation}
    \widetilde{s}_{j}^{(r)}
    =
    s_j
    \mathbin{\oplus}
    m_j^{(r)},
    \label{eq:833_noisy_syndrome_bit}
\end{equation}
where $m_j^{(r)}$ is an independent Bernoulli random variable with
\begin{equation}
    \Pr
    \left[
        m_j^{(r)}=1
    \right]
    =
    p_{\mathrm{meas}}.
    \label{eq:833_measurement_flip_probability}
\end{equation}
The variables $m_j^{(r)}$ are sampled independently for every
syndrome bit and every measurement round.

The syndrome is measured for $R$ rounds and decoded offline using a
bitwise majority vote. For each syndrome bit, the estimate is
\begin{equation}
    \widehat{s}_j
    =
    \begin{cases}
        1,
        &
        \displaystyle
        \sum_{r=1}^{R}
        \widetilde{s}_{j}^{(r)}
        >
        \frac{R}{2},
        \\[0.7em]
        0,
        &
        \displaystyle
        \sum_{r=1}^{R}
        \widetilde{s}_{j}^{(r)}
        \leq
        \frac{R}{2}.
    \end{cases}
    \label{eq:833_majority_vote}
\end{equation}
The estimated syndrome
$\widehat{\boldsymbol{s}}
=(\widehat{s}_1,\ldots,\widehat{s}_5)$ is then passed to the
weight-one decoder in Eq.~\eqref{eq:833_weight_one_decoder}. Although
Eq.~\eqref{eq:833_majority_vote} formally resolves ties in favor of
zero, all values of $R$ considered here are odd.

For odd $R$, the probability that majority voting returns an
incorrect value for one syndrome bit is
\begin{equation}
    p_{\mathrm{maj}}
    =
    \sum_{j=(R+1)/2}^{R}
    \binom{R}{j}
    p_{\mathrm{meas}}^{j}
    \left(
        1-p_{\mathrm{meas}}
    \right)^{R-j}.
    \label{eq:833_majority_error_probability}
\end{equation}
Repeated measurements therefore suppress independent syndrome-bit
errors without changing the initially applied data-error pattern.

\begin{figure}[h]
    \centering
    \includegraphics[
        width=\columnwidth
    ]{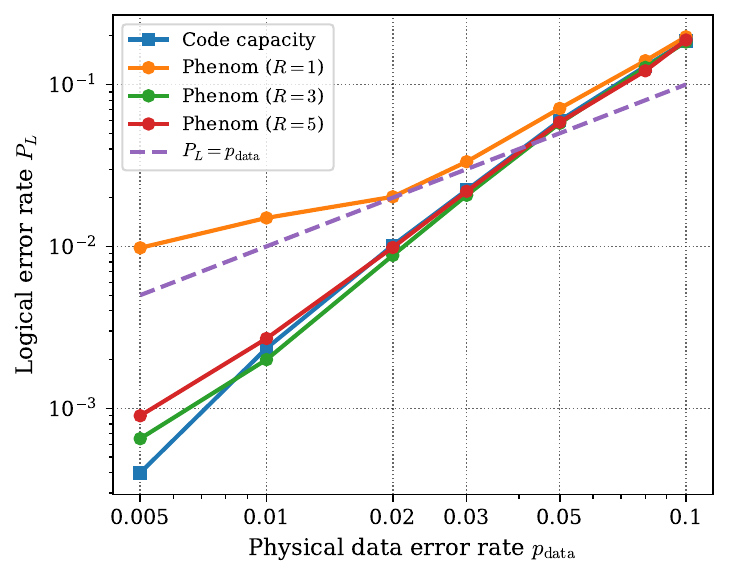}
    \caption{
        Logical error rate of the $\llbr 8,3,3\rrbr$ memory under
        phenomenological measurement noise for $R=1$, $3$, and $5$
        repeated syndrome measurements. A data error with probability
        $p_{\mathrm{data}}$ is applied once at the beginning of the
        experiment, while each reported syndrome bit is independently
        flipped with probability $p_{\mathrm{meas}}=0.005$ in every
        round. The syndrome is estimated by bitwise majority vote and
        decoded using the weight-one lookup table. The $R=3$ and
        $R=5$ results are comparable and substantially improve upon
        $R=1$. The code-capacity curve and the unencoded reference
        $P_L=p_{\mathrm{data}}$ are included for comparison. Each point is estimated from \(N=20,000\) Monte Carlo shots.
    }
    \label{fig:833_phenom_noise}
\end{figure}

Figure~\ref{fig:833_phenom_noise} compares $R=1$, $3$, and $5$ at
the fixed measurement-error probability
$p_{\mathrm{meas}}=0.005$. With a single measurement round, an
isolated syndrome-bit error is passed directly to the lookup-table
decoder, producing a substantially higher logical error rate at small
$p_{\mathrm{data}}$. Majority voting strongly suppresses this
contribution for both $R=3$ and $R=5$.

The sampled $R=3$ curve lies slightly below the $R=5$ curve over part
of the simulated parameter range. The two choices nevertheless
exhibit comparable overall logical performance. We therefore use
$R=3$ in the subsequent phenomenological-noise experiments because it
provides the lowest sampled logical error rates while requiring fewer
syndrome measurements than $R=5$. The weaker sampled performance of
$R=5$ does not imply that additional ideal repetitions intrinsically
increase the majority-vote error probability; rather, it reflects the
combined finite-sample behavior of the simulated decoder over the
parameter range considered.

\begin{figure}[h]
    \centering
    \includegraphics[
        width=\columnwidth
    ]{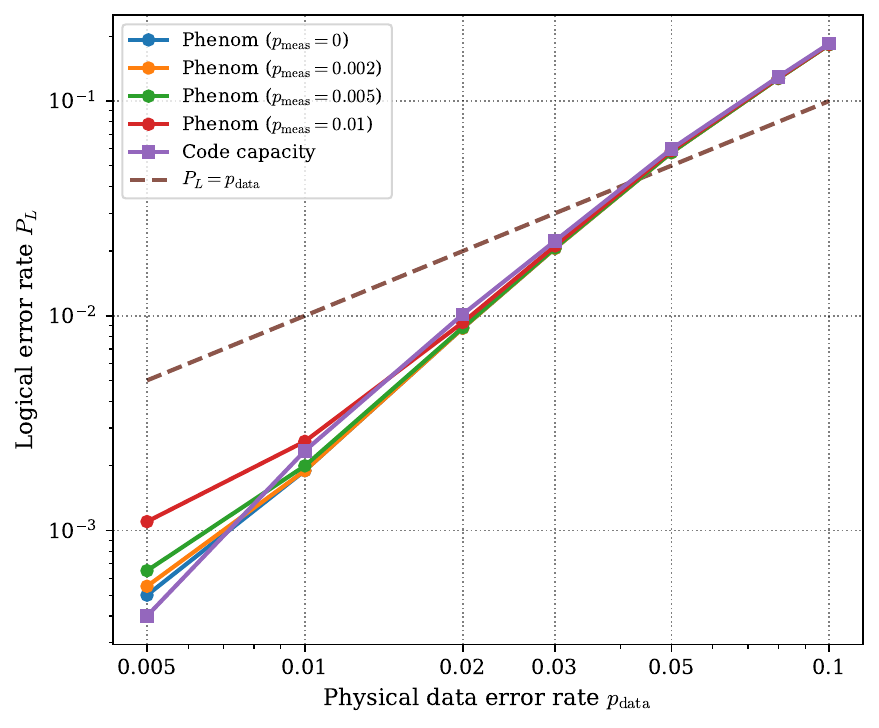}
    \caption{
        Logical error rate of the $\llbr 8,3,3\rrbr$ memory with
        $R=3$ repeated syndrome measurements. The measurement-error
        probability is varied over
        $p_{\mathrm{meas}}\in\{0,0.002,0.005,0.01\}$. A data error is
        applied once at the beginning of each experiment, and the
        syndrome is estimated using bitwise majority vote. Measurement
        noise has its strongest relative effect at small
        $p_{\mathrm{data}}$, while data errors dominate at larger
        $p_{\mathrm{data}}$. The code-capacity curve and the unencoded
        reference $P_L=p_{\mathrm{data}}$ are included for comparison. Each point is estimated from \(N=20,000\) Monte Carlo shots.
    }
    \label{fig:833_phenom_R3_pmeas}
\end{figure}

Motivated by this comparison, we fix $R=3$ and independently vary the
measurement-error probability over
$p_{\mathrm{meas}}\in\{0,0.002,0.005,0.01\}$. The resulting logical
error rates are shown in Fig.~\ref{fig:833_phenom_R3_pmeas}. The
$p_{\mathrm{meas}}=0$ curve provides a reference for isolating the
additional contribution from imperfect syndrome measurements.

For nonzero $p_{\mathrm{meas}}$, the degradation is most visible at
small $p_{\mathrm{data}}$, where syndrome-readout errors constitute a
relatively large fraction of the observed failures. At larger
$p_{\mathrm{data}}$, the curves approach one another as data errors
become the dominant source of logical failure.

Together, the code-capacity and phenomenological-noise experiments
validate the weight-one lookup-table decoder and recover the expected
distance-three suppression of single-qubit data errors. They also
show that three rounds of majority-vote syndrome estimation
sufficiently suppress the measurement-error contribution over the
parameter range considered here. We next replace ideal syndrome
readout by explicit noisy extraction circuits.

\subsection{Circuit-level error correction and flagged syndrome extraction}
\label{sec:833_flagged_qec}

The preceding memory experiments treat syndrome readout at the
phenomenological level and do not model faults within the extraction
circuits. We now introduce an explicit circuit-level error-correction
procedure based on the two-ancilla flag construction of
Ref.~\cite{chao2018quantum}. The syndrome ancilla extracts a
stabilizer eigenvalue, while an additional flag ancilla detects faults
that could otherwise propagate through the circuit and produce
correlated errors on the data qubits.

For the flagged construction, we replace the first two generators in
Table~\ref{tab:833_code} by their products with $S_3$ and use the
equivalent generating set
\begin{equation}
    \mathcal{S}_{833}
    =
    \left\langle
        g_1,g_2,g_3,g_4,g_5
    \right\rangle ,
    \label{eq:833_extraction_generating_set}
\end{equation}
where
\begin{align}
    g_1 &= S_1S_3
    = X_1X_2Y_3Z_4Y_6Z_7,
    \nonumber\\
    g_2 &= S_2S_3
    = Z_1Z_2X_4Y_5X_7Y_8,
    \nonumber\\
    g_3 &= S_3
    = Z_3Y_4X_5Z_6Y_7X_8,
    \nonumber\\
    g_4 &= S_4
    = Z_2X_3X_5Y_6Z_7Y_8,
    \nonumber\\
    g_5 &= S_5
    = X_2Z_4Z_5X_6Y_7Y_8.
    \label{eq:833_extraction_generators}
\end{align}
This replacement does not change the stabilizer group or the
codespace. It reduces the weights of the first two generators from
eight to six, so that all five extraction generators have equal
weight and admit circuits of comparable depth.

More importantly, the choice of extraction generators and the order
of their controlled-Pauli interactions are designed together. Within
a flagged extraction circuit, a syndrome-ancilla fault may propagate
through the remaining controlled-Pauli gates and produce a correlated
Pauli error on several data qubits. For fault-tolerant recovery, the
possible correlated errors associated with a triggered flag must be
distinguishable from the complete syndrome obtained afterward (assumed to be ideal as the single-fault budget has already been met). The
equivalent generating set in
Eq.~\eqref{eq:833_extraction_generators}, together with the
generator-dependent gate ordering shown below, ensures that the
relevant propagated errors have distinct full (ideal) syndromes within each
flagged circuit.

For a Pauli error $E$, the syndrome used throughout the numerical
experiments is therefore
\begin{equation}
    \boldsymbol{s}(E)
    =
    \bigl(
        s_1(E),s_2(E),s_3(E),s_4(E),s_5(E)
    \bigr),
    \label{eq:833_error_syndrome}
\end{equation}
where $s_i(E)=0$ if $E$ commutes with $g_i$, and $s_i(E)=1$ if
$E$ anticommutes with $g_i$. The syndrome bits are ordered according
to the extraction circuits for $(g_1,g_2,g_3,g_4,g_5)$. This is the
same convention used by the weight-one decoder in Sec.~\ref{sec:833_memory}. The complete weight-one syndrome map is
listed in Table~\ref{tab:833_weight1_lut}, while the
circuit-dependent flag-conditioned maps are listed in
Table~\ref{tab:833_flag_luts}.

\begin{figure*}[t]
    \centering

    \begin{tabular}{@{}ccc@{}}
        \includegraphics[width=0.30\textwidth]
        {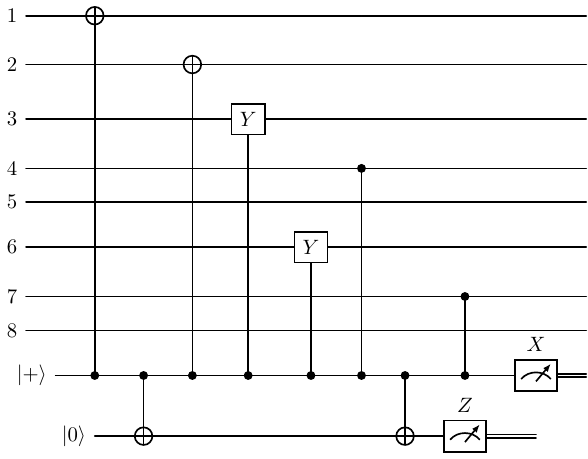}
        &
        \includegraphics[width=0.30\textwidth]
        {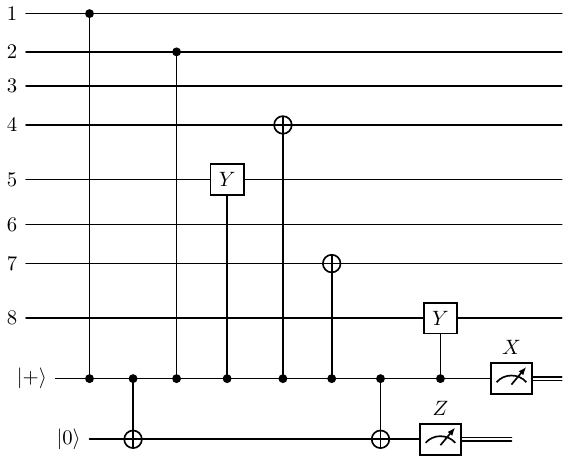}
        &
        \includegraphics[width=0.30\textwidth]
        {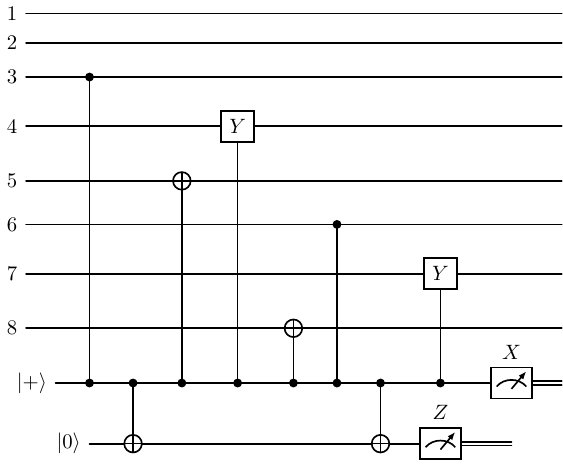}
        \\
        \textnormal{(a) Flagged extraction of $g_1$.}
        &
        \textnormal{(b) Flagged extraction of $g_2$.}
        &
        \textnormal{(c) Flagged extraction of $g_3$.}
    \end{tabular}

    \vspace{1.0em}

    \begin{tabular}{@{}cc@{}}
        \includegraphics[width=0.30\textwidth]
        {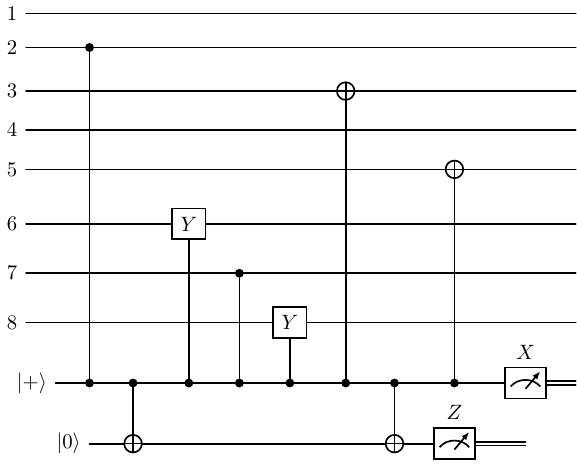}
        &
        \includegraphics[width=0.30\textwidth]
        {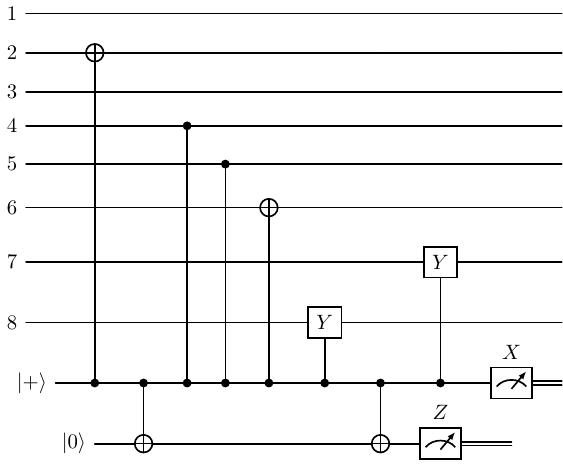}
        \\
        \textnormal{(d) Flagged extraction of $g_4$.}
        &
        \textnormal{(e) Flagged extraction of $g_5$.}
    \end{tabular}

    \caption{
        Flagged syndrome-extraction circuits for the five generators
        in Eq.~\eqref{eq:833_extraction_generators}. The upper eight
        wires represent the data qubits. In each panel, the syndrome
        ancilla is prepared in $\ket{+}$ and measured in the $X$
        basis, while the flag ancilla is prepared in $\ket{0}$ and
        measured in the $Z$ basis. The controlled-Pauli interactions
        are ordered so that the correlated errors associated with a
        triggered flag have distinct full syndromes within each
        extraction circuit.
    }
    \label{fig:833_flagged_circuits}
\end{figure*}

The five flagged syndrome-extraction circuits are shown in
Fig.~\ref{fig:833_flagged_circuits}. In each circuit, the syndrome
ancilla is prepared in $\ket{+}$ and measured in the $X$ basis,
whereas the flag ancilla is prepared in $\ket{0}$ and measured in the
$Z$ basis. The controlled-Pauli interactions between the syndrome
ancilla and the data qubits encode the eigenvalue of $g_i$ in the
syndrome-ancilla measurement. The two interactions between the
syndrome and flag ancillas are positioned such that a
syndrome-ancilla fault capable of producing a relevant correlated
data error also produces a nontrivial flag outcome.

Errors arising from different flagged circuits need not have globally
distinct syndromes. The identity of the circuit in which the flag was
triggered is retained as part of the decoder input. It is therefore
sufficient for the relevant correlated errors to have distinct
syndromes within each individual extraction circuit.

The error-correction procedure processes the generators sequentially
in the order $g_1,\ldots,g_5$. For each $g_i$, the corresponding
flagged circuit in Fig.~\ref{fig:833_flagged_circuits} is executed. If
both the syndrome and flag outcomes are trivial, the protocol
continues to $g_{i+1}$. If either outcome is nontrivial, all five
generators are subsequently measured using unflagged extraction
circuits to obtain the complete syndrome
$\boldsymbol{s}=(s_1,\ldots,s_5)$. Again, this is assumed to be ideal
since the single-fault budget of the distance-$3$ code has already been
met. In simulations, though, these unflagged circuits are still noisy; 
if the complete syndrome is not recognized in the table due to any 
additional error during unflagged extraction, then the round is discarded
and not counted towards the logical error rate.

The unflagged circuit for $g_i$ is obtained from its flagged
counterpart by removing the flag ancilla and the two interactions
between the syndrome and flag ancillas. The syndrome ancilla and all
controlled-Pauli interactions with the data qubits are retained.
Since the unflagged circuits follow directly from
Fig.~\ref{fig:833_flagged_circuits}, they are not displayed
separately.

If the flag is triggered during the extraction of $g_i$, the pair
$(i,\boldsymbol{s})$ is decoded using the circuit-dependent map
\begin{equation}
    \mathcal{D}_{\mathrm{flag}}^{(i)}
    \bigl(\boldsymbol{s}\bigr)
    =
    E_{\boldsymbol{s}}^{(i)},
    \label{eq:833_flag_decoder}
\end{equation}
where $E_{\boldsymbol{s}}^{(i)}$ is the correlated Pauli error
associated with (ideal) syndrome $\boldsymbol{s}$ in the extraction circuit
for $g_i$. The corresponding flag-conditioned lookup tables are
listed in Table~\ref{tab:833_flag_luts}. Since Pauli operators are
self-inverse up to a physically irrelevant global phase, the inferred
error itself may be applied as the recovery operator.

If the measured syndrome is nontrivial but the flag is not triggered,
then the complete unflagged syndrome is decoded using the weight-one lookup table,
\begin{equation}
    E_{\mathrm{rec}}
    =
    \mathcal{D}_{\mathrm{LUT}}
    \bigl(\boldsymbol{s}\bigr).
    \label{eq:833_unflagged_recovery}
\end{equation}
The complete map $\mathcal{D}_{\mathrm{LUT}}$ is given in Table~\ref{tab:833_weight1_lut}. The weight-one and flag-conditioned lookup tables use the same syndrome ordering $(g_1,g_2,g_3,g_4,g_5)$, so no syndrome-basis conversion is required.

We simulate this code-specific implementation of the
Chao--Reichardt protocol using the Stim stabilizer-circuit
simulator~\cite{chao2018quantum,gidney2021stim}. In the circuit-level
noise model, every single-qubit gate is followed by a single-qubit
depolarizing channel with probability $p$, and every two-qubit gate
is followed by a two-qubit depolarizing channel with the same
probability $p$. The encoded input states are prepared through ideal
projective measurements of the stabilizer generators and the
appropriate logical Pauli operators, as described below. Ancilla
reset, projective measurements, and final logical readout are treated
as ideal unless stated otherwise.

The encoded input state is prepared projectively. Starting from a
physical product state, we perform ideal projective measurements of
the five extraction generators \(g_1,\ldots,g_5\). The measurements
project the state into a definite stabilizer-syndrome sector. For a
nontrivial syndrome contained in the weight-one lookup table in Table~\ref{tab:833_weight1_lut}, the recovery prescribed by \(\mathcal{D}_{\mathrm{LUT}}\) is applied, thereby mapping the state to the simultaneous \(+1\) eigenspace of the stabilizer group. The trivial syndrome requires no recovery. If the measured syndrome lies outside the weight-one lookup table, the preparation attempt is rejected and restarted. Since state preparation is ideal and occurs before the noisy circuit begins, this rejection affects only the preparation cost and is not included in the reported logical failure rate.

For the logical-$Z$ ensemble, ideal projective measurements of
$\overline{Z_1}$, $\overline{Z_2}$, and $\overline{Z_3}$ are
performed. Whenever the measurement of $\overline{Z_j}$ returns
eigenvalue $-1$, the corresponding logical operator
$\overline{X_j}$ is applied to map it to the $+1$ eigenspace. This
prepares $\ket{000}_{L}$. Similarly, the logical-$X$ ensemble is
prepared by projectively measuring $\overline{X_1}$,
$\overline{X_2}$, and $\overline{X_3}$ and applying
$\overline{Z_j}$ following any $-1$ outcome, thereby preparing
$\ket{+++}_{L}$.

We first use the flagged dynamic error-correction procedure as a
circuit-level memory experiment. Following the Chao--Reichardt
protocol~\cite{chao2018quantum}, the extraction circuits are processed
sequentially in the order $g_1,\ldots,g_5$. For each $g_i$, the
corresponding flagged circuit in
Fig.~\ref{fig:833_flagged_circuits} is executed. If both the syndrome
and flag outcomes are trivial, the protocol continues to $g_{i+1}$.
If either outcome is nontrivial, all five generators are measured
using unflagged extraction circuits, and the resulting complete
syndrome is decoded using either the flag-conditioned or weight-one
lookup table, as described above.

We consider $R$ consecutive executions of the complete
error-correction procedure, where $R$ denotes the number of full
flagged error-correction rounds. Among the sampled choices, $R=1$
gives the lowest logical error rates. We therefore use a single round
as the circuit-level memory baseline and throughout the subsequent
plain-Trotter simulations.

At the end of a noisy error-correction round, a correctable
weight-one Pauli error may remain on the data block. Although such an
error has not produced a logical failure, it may anticommute with a
physical representative of a logical observable and thereby flip a
direct final measurement. Counting this outcome as a logical failure
would conflate a correctable residual data error with an uncorrectable
logical error.

We therefore apply an ideal terminal error-correction step before the
final logical readout. Specifically, we perform an ideal full
syndrome-extraction pass and apply the corresponding weight-one
recovery from $\mathcal{D}_{\mathrm{LUT}}$, as listed in
Table~\ref{tab:833_weight1_lut}. This terminal cleanup is
not part of the noisy error-correction protocol. It removes
correctable residual errors so that the reported logical failure rate
measures the logical content of the final state after ideal decoding.
The same terminal convention is used in all subsequent circuit-level
simulations.

For the $\ket{000}_{L}$ ensemble, we measure
$\overline{Z_1}$, $\overline{Z_2}$, and $\overline{Z_3}$ after the
ideal cleanup. A change in any of their eigenvalues is counted as a
logical-$X$ failure. For the $\ket{+++}_{L}$ ensemble, we instead
measure $\overline{X_1}$, $\overline{X_2}$, and
$\overline{X_3}$, and a change in any of their eigenvalues is counted
as a logical-$Z$ failure. Accordingly, we define
\begin{align}
    P_{L,X}
    &=
    \Pr
    \left[
        \text{at least one }
        \overline{Z_j}
        \text{ eigenvalue changes}
    \right],
    \nonumber\\
    P_{L,Z}
    &=
    \Pr
    \left[
        \text{at least one }
        \overline{X_j}
        \text{ eigenvalue changes}
    \right].
    \label{eq:833_circuit_logical_error_rates}
\end{align}

\begin{figure}[h]
    \centering
    \includegraphics[
        width=\columnwidth
    ]{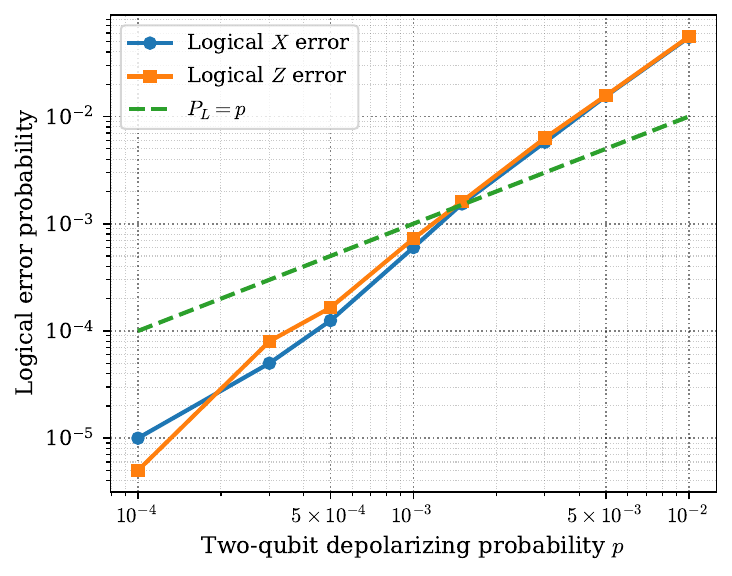}
    \caption{
        Circuit-level logical error probabilities of the
        $\llbr 8,3,3\rrbr$ memory under one round of the
        Chao--Reichardt flagged error-correction
        protocol~\cite{chao2018quantum}. Single- and two-qubit gates are followed by the corresponding single- and two-qubit depolarizing channels with probability \(p\). An ideal weight-one cleanup is performed before the final logical measurements. Each point is estimated from $N=200,000$ Monte Carlo shots. The logical-$X$ and logical-$Z$ curves cross the unencoded reference $P_L=p$ near $p\simeq1.5\times10^{-3}$.
    }
    \label{fig:833_rui_memory}
\end{figure}

The circuit-level memory results are shown in
Fig.~\ref{fig:833_rui_memory}. Each data point in this figure is estimated from
$N=200,000$ Monte Carlo shots. Both logical-error curves cross the
unencoded reference $P_L=p$ near
$p\simeq1.5\times10^{-3}$, giving a circuit-level pseudo-threshold of
approximately $1.5\times10^{-3}$ for the single-round protocol. Shot counts for the remaining numerical experiments are reported in
their respective figure captions.

\begin{remark}
For comparison, we also simulate the protocol without the final ideal
cleanup. In this case, residual weight-one errors produced near the
end of the noisy circuit remain on the data block at readout, and no
pseudo-threshold is observed over the sampled parameter range. This
comparison shows that the treatment of correctable residual errors at
the final circuit boundary materially affects the reported
finite-circuit logical error rate. To maintain a consistent boundary
convention, all subsequent circuit-level simulations in this work
include the same final ideal weight-one cleanup.
\end{remark}

\subsection{Encoded Trotter Implementation}

Next, we insert an encoded logical Trotter circuit between state
preparation and the final error-correction stage. As shown in our
previous work on fault-tolerant quantum simulation using symplectic
transvections~\cite{chen2025fault}, the encoded physical circuit can
preserve the Trotter pattern of the corresponding unencoded
evolution. This correspondence allows the same product-formula
structure to be implemented directly on an encoded data block.

\begin{figure}[h]
    \centering
    \includegraphics[
        width=\columnwidth
    ]{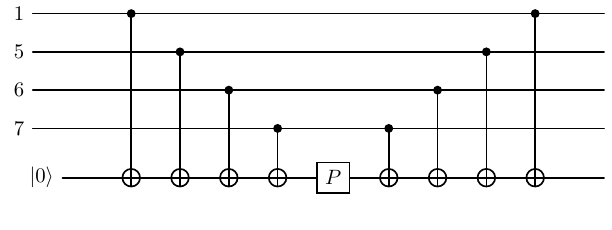}
    \caption{
        Physical implementation of the encoded Pauli rotation in
        Eq.~\eqref{eq:833_encoded_trotter_unitary}. The auxiliary
        qubit accumulates the parity of data qubits $1$, $5$, $6$,
        and $7$, which support
        $\overline{Z_2}=Z_1Z_5Z_6Z_7$. The central phase gate
        implements the $\pi/4$ rotation, and the remaining CNOT gates
        uncompute the parity.
    }
    \label{fig:833_plain_trotter_circuit}
\end{figure}

As a baseline example, we consider the single-qubit (logical) Pauli rotation
\begin{equation}
    U
    =
    \exp
    \left(
        -i\frac{\pi}{4}Z_2
    \right).
    \label{eq:833_unencoded_trotter_unitary}
\end{equation}
For the logical representation in Table~\ref{tab:833_code},
\begin{equation}
    \overline{Z_2}
    =
    Z_1Z_5Z_6Z_7.
    \label{eq:833_trotter_logical_pauli}
\end{equation}
The corresponding encoded evolution is
\begin{equation}
    \overline{U}
    =
    \exp
    \left(
        -i\frac{\pi}{4}\overline{Z_2}
    \right)
    =
    \exp
    \left(
        -i\frac{\pi}{4}Z_1Z_5Z_6Z_7
    \right).
    \label{eq:833_encoded_trotter_unitary}
\end{equation}
Its physical implementation is shown in
Fig.~\ref{fig:833_plain_trotter_circuit}. 

%
\begin{figure}[t]
    \centering
    \includegraphics[
        width=\columnwidth
    ]{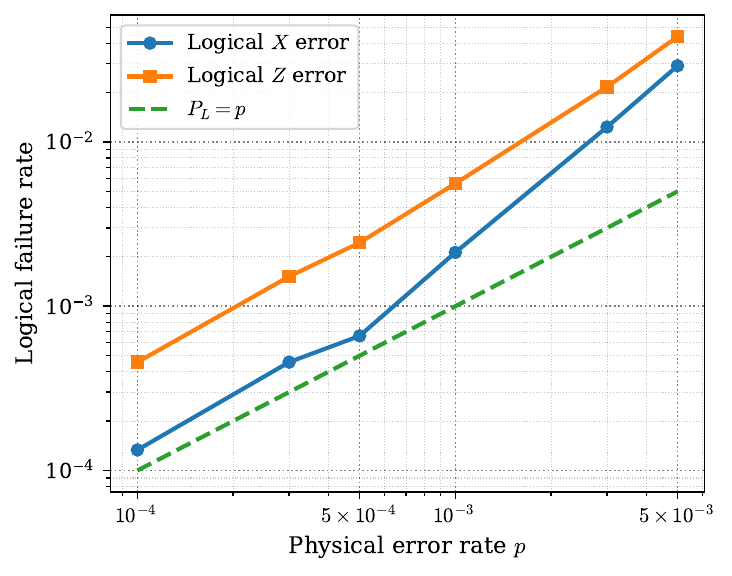}
    \caption{
        Logical-$X$ and logical-$Z$ failure rates for the encoded
        Trotter evolution in
        Eq.~\eqref{eq:833_encoded_trotter_unitary}, followed by one
        round of Chao--Reichardt flagged error
        correction~\cite{chao2018quantum} and a final ideal
        weight-one cleanup. The final logical measurements use the
        ideally propagated observables rather than the corresponding
        bare input observables. Neither logical-error curve exhibits
        a pseudo-threshold relative to the unencoded reference
        $P_L=p$ over the sampled parameter range. Each point is estimated from \(N=200,000\) Monte Carlo shots.
    }
    \label{fig:833_plain_trotter}
\end{figure}

The logical failure rates of the plain encoded-Trotter implementation
are shown in Fig.~\ref{fig:833_plain_trotter}.
The logical-$X$ and logical-$Z$ failure rates are evaluated using
separate input ensembles. For the logical-$Z$ ensemble, the input
state is projected into the simultaneous $+1$ eigenspace of
$\overline{Z_1}$, $\overline{Z_2}$, and $\overline{Z_3}$. Logical-$X$
failures are detected through the corresponding final logical-$Z$
observables. For the logical-$X$ ensemble, the input state is
projected into the simultaneous $+1$ eigenspace of
$\overline{X_1}$, $\overline{X_2}$, and $\overline{X_3}$, and
logical-$Z$ failures are detected through the final logical-$X$
observables.
Since the intended logical Trotter evolution can transform the
measured Pauli observables, the final measurements are not performed
using the bare input observables. For each initial logical observable
$\overline{M}$, we measure the ideally propagated observable
$\overline{M_{\mathrm{out}}}
=
\overline{U}\mkern4mu
\overline{M}\mkern4mu
\overline{U}^{\dagger}$.
A logical failure is recorded when the measured eigenvalue differs
from the value predicted by the ideal logical evolution. Thus, the
intended transformation of a logical observable is not itself
classified as an error.

After the noisy encoded Trotter circuit and the flagged
error-correction round, we apply the same final ideal cleanup used in
the circuit-level memory experiment. In contrast with the memory
benchmark in Fig.~\ref{fig:833_rui_memory}, neither logical-error
curve in Fig.~\ref{fig:833_plain_trotter} crosses below the unencoded
reference over the sampled parameter range. Thus, no pseudo-threshold
is observed for the plain encoded-Trotter implementation. The marked
asymmetry between the two logical-error channels can be understood
from error propagation through the parity-uncomputation circuit.

Let $a$ denote the auxiliary qubit. For a CNOT with data qubit $j$ as
the control and $a$ as the target, a $Z$ error on the target
propagates according to
\begin{equation}
    \operatorname{CNOT}_{j\rightarrow a}
    Z_a
    \operatorname{CNOT}_{j\rightarrow a}
    =
    Z_jZ_a.
    \label{eq:833_target_z_propagation}
\end{equation}
Consequently, a single $Z_a$ error occurring immediately after the
phase gate propagates through the four CNOT gates in the
parity-uncomputation circuit as
\begin{equation}
    Z_a
    \longmapsto
    Z_1Z_5Z_6Z_7Z_a.
    \label{eq:833_trotter_target_z_propagation}
\end{equation}
The auxiliary qubit is not part of the encoded data block, so its
residual $Z_a$ component is not included in the logical-error
classification. Restricting the propagated error in
Eq.~\eqref{eq:833_trotter_target_z_propagation} to the data qubits
gives
\begin{equation}
    \left.
        Z_1Z_5Z_6Z_7Z_a
    \right|_{\mathrm{data}}
    =
    Z_1Z_5Z_6Z_7
    =
    \overline{Z_2}.
    \label{eq:833_trotter_logical_z_propagation}
\end{equation}
Thus, a single fault on the auxiliary qubit produces the complete
logical operator $\overline{Z_2}$ on the data block. Since a logical
Pauli operator has trivial stabilizer syndrome, neither the
subsequent flagged error-correction round nor the final weight-one
cleanup can remove this error. The plain Trotter gadget therefore has
circuit-level distance one in the logical-$Z$ error sector. This
direct propagation mechanism explains the substantially larger
logical-$Z$ failure rate observed in
Fig.~\ref{fig:833_plain_trotter}.

The absence of a logical-$X$ pseudo-threshold does not admit an
equally simple target-error propagation path. Nevertheless, the
approximately linear low-noise behavior of the logical-$X$ curve is
consistent with first-order malignant fault contributions. In
particular, a two-qubit Pauli fault in the parity-computation circuit
can introduce errors on both a data qubit and the auxiliary qubit.
The auxiliary component may subsequently be transformed by the phase
gate and propagated through the parity-uncomputation circuit,
producing a correlated data error that lies outside the weight-one
lookup table.

More generally, the encoded implementation contains substantially
more noisy circuit locations than the corresponding unencoded Pauli
rotation. Its low-noise logical-$X$ failure rate can therefore contain
a leading contribution of the form
\begin{equation}
    P_{L,X}
    =
    A_Xp
    +
    O(p^2),
    \label{eq:833_plain_trotter_lx_scaling}
\end{equation}
where $A_X$ collects the contributions from malignant single-fault
locations, which encompasses a failed CNOT gate causing a two-qubit error. 
The numerical results are consistent with $A_X>1$ over
the sampled range, preventing the logical-$X$ curve from crossing
below the unencoded reference. A complete identification of
the contributing locations requires an explicit circuit-level
single-fault enumeration. These observations establish the plain encoded-Trotter circuit as the baseline for the circuit-level fault-mitigation strategies considered below.

\subsection{Addressing Problematic Single-Faults}

\subsubsection{Biased Noise with Flags}

The absence of a pseudo-threshold in
Fig.~\ref{fig:833_plain_trotter} motivates a noise model tailored to
the dominant fault-propagation mechanism of the encoded Trotter
circuit. In particular, the propagation mechanism in
Eq.~\eqref{eq:833_trotter_target_z_propagation} originates from phase
faults on the auxiliary parity qubit. We therefore first consider a
restricted $Z$-biased noise model for the Trotter block. This model is
motivated by stabilized cat-qubit architectures, in which bit-flip
errors can be exponentially suppressed while native two-qubit gates
preserve a noise channel dominated by phase-flip
errors~\cite{puri2020bias}.

For a CNOT location in the Trotter circuit, the fully-$Z$-biased two-qubit
channel is defined as
\begin{equation}
    \mathcal{N}^{(2)}_{Z,p}(\rho)
    =
    (1-p)\rho
    +
    \frac{p}{3}
    \sum_{E\in\{IZ,ZI,ZZ\}}
    E\rho E,
    \label{eq:833_z_biased_two_qubit_channel}
\end{equation}
whereas a phase-gate location is followed by
\begin{equation}
    \mathcal{N}^{(1)}_{Z,p}(\rho)
    =
    (1-p)\rho
    +
    pZ\rho Z.
    \label{eq:833_z_biased_single_qubit_channel}
\end{equation}
The restriction in
Eqs.~\eqref{eq:833_z_biased_two_qubit_channel}
and~\eqref{eq:833_z_biased_single_qubit_channel} is applied only to
the encoded Trotter block. The subsequent syndrome-extraction
circuits contain CNOT, CZ, and CY interactions and retain the
depolarizing circuit-level noise model used in the memory benchmark.
Thus, Eqs.~\eqref{eq:833_z_biased_two_qubit_channel}
and~\eqref{eq:833_z_biased_single_qubit_channel} should be interpreted
as an idealized biased-noise limit for the Trotter operation rather
than as a complete microscopic noise model for the entire experiment.

To obtain additional information about phase-error propagation during
the encoded Trotter operation, we augment the circuit with two
$Z$-error-sensitive flag ancillas, hereafter referred to as $Z$-type
flags. Each flag ancilla is prepared in $\ket{+}$, coupled twice to
the auxiliary parity qubit as a CNOT control, and measured in the
$X$ basis. A $Z$ error that appears on the parity qubit between the
two flag couplings propagates a $Z$ component onto the corresponding
flag ancilla and can therefore change its $X$-basis measurement
outcome.

\begin{figure*}[t]
    \centering
    \includegraphics[
        width=0.90\textwidth
    ]{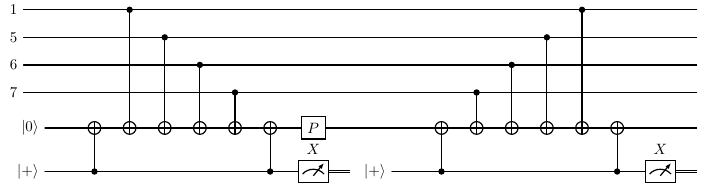}
    \caption{
        Encoded Trotter circuit equipped with two $Z$-type flag
        ancillas. The left flag monitors phase-error propagation
        during parity computation, while the right flag monitors
        parity uncomputation. Each flag is prepared in $\ket{+}$,
        coupled twice to the auxiliary parity qubit, and measured in
        the $X$ basis. Error classification is performed on the
        eight-qubit data block; Pauli components acting only on the
        auxiliary parity qubit are not counted as data errors.
    }
    \label{fig:833_flagged_trotter_circuit}
\end{figure*}

The two $Z$-type flags in
Fig.~\ref{fig:833_flagged_trotter_circuit} monitor different portions
of the Trotter circuit. The left flag monitors the parity-computation
portion before the phase gate, whereas the right flag monitors the
parity-uncomputation portion after the phase gate. We denote their
binary outcomes by $f_{\mathrm L}$ and $f_{\mathrm R}$, respectively,
where an outcome of $1$ indicates a triggered flag:
\begin{equation}
    \left(
        f_{\mathrm L},
        f_{\mathrm R}
    \right)
    \in
    \left\{
        (0,0),(1,0),(0,1),(1,1)
    \right\}.
    \label{eq:833_z_flag_outcomes}
\end{equation}
The auxiliary parity qubit is not part of the encoded data block.
Consequently, Pauli components supported only on this auxiliary qubit
are omitted when classifying the residual data error.

Under the restricted noise model, propagation of the candidate
single-location faults shows that the nontrivial $Z$-type data errors
associated with a single triggered flag are contained in
\begin{equation}
    \mathcal{E}_{Z,\mathrm{flag}}
    =
    \left\{
        Z_1,\,
        Z_1Z_5,\,
        Z_1Z_5Z_6,\,
        Z_1Z_5Z_6Z_7
    \right\}.
    \label{eq:833_flagged_trotter_error_set}
\end{equation}
The corresponding flag-conditioned lookup table is shown in
Table~\ref{tab:833_z_biased_trotter_lut}.
\begin{table}[h]
    \centering
    \caption{
        Syndrome-conditioned recovery table used after exactly one
        $Z$-type flag is triggered in the biased-noise encoded-Trotter
        experiment. Syndrome bits are ordered according to the five
        syndrome-extraction circuits.
    }
    \label{tab:833_z_biased_trotter_lut}
    \begin{tabular}{cc}
        \toprule
        Syndrome & Recovery \\
        \midrule
        $10000$ & $Z_1$ \\
        $11110$ & $Z_1Z_5$ \\
        $01101$ & $Z_1Z_5Z_6$ \\
        $00000$ & $Z_1Z_5Z_6Z_7$ \\
        \bottomrule
    \end{tabular}
\end{table}

The recovery procedure is conditioned on the two $Z$-flag outcomes.
If neither flag is triggered, corresponding to
$(f_{\mathrm L},f_{\mathrm R})=(0,0)$, we apply the standard noisy
Chao--Reichardt error-correction procedure. If exactly one flag is
triggered, so that
\begin{equation}
    \left(
        f_{\mathrm L},
        f_{\mathrm R}
    \right)
    \in
    \left\{
        (1,0),(0,1)
    \right\},
    \label{eq:833_single_z_flag_outcomes}
\end{equation}
we perform one noisy full syndrome-extraction pass using the five
unflagged extraction circuits. The resulting five-bit syndrome is
decoded using Table~\ref{tab:833_z_biased_trotter_lut}. Runs in which
both flags are triggered are discarded. Finally, every accepted shot
is followed by the same ideal cleanup procedure used in the preceding
circuit-level simulations.

\begin{figure}[h]
    \centering
    \includegraphics[
        width=\columnwidth
    ]{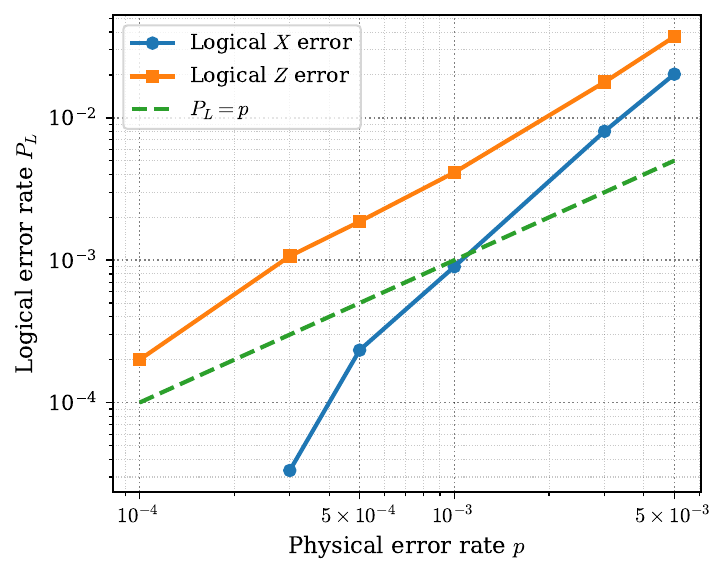}
    \caption{
        Logical error rates for the flagged encoded-Trotter circuit
        under the restricted $Z$-biased Trotter noise model. If
        exactly one $Z$-type flag is triggered, the recovery rule in
        Table~\ref{tab:833_z_biased_trotter_lut} is applied after a
        noisy unflagged syndrome-extraction pass. Double-flag events
        are discarded, and an ideal cleanup pass is applied before
        the final propagated logical measurements. A pseudo-threshold
        is observed in the logical-$X$ channel, but not in the
        logical-$Z$ channel. Each point is estimated from \(N=200,000\) Monte Carlo shots.
    }
    \label{fig:833_z_biased_trotter}
\end{figure}

The resulting logical error rates are shown in
Fig.~\ref{fig:833_z_biased_trotter}. The $Z$-biased model together
with the flag-conditioned recovery substantially improves the
logical-$X$ performance, for which a pseudo-threshold is observed. In
contrast, the logical-$Z$ error rate remains above the unencoded
reference over the sampled physical-error range. Thus, even after
suppressing the non-$Z$ components of the Trotter noise and
incorporating two $Z$-type flags, the encoded implementation does not
obtain a pseudo-threshold in both logical channels.

\subsubsection{Complete Single-Fault Analysis}

The limited logical-$Z$ improvement motivates a closer examination of
whether flag measurements can, more generally, provide sufficient
information to protect the encoded Trotter circuit. This question is
distinct from the restricted biased-noise simulation above. We
therefore perform an exhaustive circuit-level single-fault analysis
using the full Pauli fault set: after each CNOT location, we enumerate
all 15 nontrivial two-qubit Pauli faults, and after the phase gate we
enumerate the three single-qubit Pauli faults. Each fault is propagated
to the end of the circuit, where we record the residual data error,
the two flag outcomes, and the ideal five-bit stabilizer syndrome.

The complete classical record available to a lookup-table decoder is
\begin{equation}
    \left(
        f_{\mathrm L},
        f_{\mathrm R},
        \boldsymbol{s}
    \right),
    \label{eq:833_complete_flag_record}
\end{equation}
where $\boldsymbol{s}$ denotes the stabilizer syndrome. Different
physical faults may produce the same record without causing a
decoding ambiguity if their residual data errors differ only by a
stabilizer. In contrast, if two errors with the same record differ by
a nontrivial logical operator, no recovery conditioned only on
Eq.~\eqref{eq:833_complete_flag_record} can correct both.

The $Z$-type construction in
Fig.~\ref{fig:833_flagged_trotter_circuit} already contains explicit
examples of such logical-coset collisions. In particular, the record
\begin{equation}
    \left(
        f_{\mathrm L},
        f_{\mathrm R},
        \boldsymbol{s}
    \right)
    =
    \left(
        1,0,10111
    \right)
    \label{eq:833_collision_record_10111}
\end{equation}
can arise from the residual data errors
\begin{equation}
    E_a
    =
    X_5Z_6Z_7,
    \qquad
    E_b
    =
    X_7.
    \label{eq:833_collision_errors_10111}
\end{equation}
Although $E_a$ and $E_b$ have identical stabilizer syndromes and
identical flag outcomes, their product is, up to an irrelevant phase,
\begin{equation}
    E_aE_b
    \sim
    X_5Z_6Y_7.
    \label{eq:833_collision_difference_10111}
\end{equation}
This operator commutes with the stabilizer group but is not itself a
stabilizer. The two residual errors therefore belong to different
logical cosets and cannot be corrected by a common recovery.

A second collision occurs for the record
\begin{equation}
    \left(
        f_{\mathrm L},
        f_{\mathrm R},
        \boldsymbol{s}
    \right)
    =
    \left(
        1,0,01000
    \right),
    \label{eq:833_collision_record_01000}
\end{equation}
which can arise from
\begin{equation}
    E_c
    =
    Y_1Z_5Z_6Z_7,
    \qquad
    E_d
    =
    Y_6Z_7.
    \label{eq:833_collision_errors_01000}
\end{equation}
Their product satisfies
\begin{equation}
    E_cE_d
    \sim
    Y_1Z_5X_6,
    \label{eq:833_collision_difference_01000}
\end{equation}
which is also a nontrivial element of the logical normalizer rather
than a stabilizer. Consequently, the two $Z$-flag outcomes together
with the full stabilizer syndrome do not always determine a logically
consistent recovery.

The simplest ambiguity appears in the trivial-syndrome sector. For
the logical representative used here,
\begin{equation}
    \overline{Z_2}
    =
    Z_1Z_5Z_6Z_7,
    \label{eq:833_z2_logical_representative}
\end{equation}
and therefore
\begin{equation}
    \boldsymbol{s}(I)
    =
    \boldsymbol{s}\!\left(
        \overline{Z_2}
    \right)
    =
    00000.
    \label{eq:833_flagged_trotter_zero_syndrome_ambiguity}
\end{equation}
The single-fault enumeration contains records for which the same flag
pattern and the trivial syndrome are compatible with either $I$ or
$\overline{Z_2}$. Mapping this record to $\overline{Z_2}$ corrects
one class of faults but introduces a logical operator when the
residual data error is the identity. Conversely, mapping it to the
identity leaves an existing $\overline{Z_2}$ error uncorrected. This
ambiguity cannot be removed by enlarging a lookup table because
stabilizer measurements cannot distinguish errors that differ by a
logical operator.

The fault analysis also exposes limitations that are not captured by
syndrome collisions alone. Some benign faults, including faults whose
restriction to the data block is the identity, trigger a flag. At the
same time, some correlated and logical data errors leave both flags
untriggered. Hence, a triggered flag is not in one-to-one
correspondence with dangerous error propagation. In a noisy circuit,
these two behaviors respectively generate false-positive flag events
and missed detections.

The two flags in Fig.~\ref{fig:833_flagged_trotter_circuit} are
specifically designed to monitor $Z$-type propagation on the
auxiliary parity qubit. To determine whether the difficulty is caused
only by this choice of flag observable, we also examined a
complementary $X$-sensitive flag construction. However, monitoring the
additional propagation channels requires several extra entangling
gates around the Trotter block. These gates increase the circuit depth
and create additional fault locations, while the resulting flag and
syndrome information still does not uniquely determine every residual
logical coset. The additional circuit overhead therefore does not
resolve the underlying decoding ambiguity.

The fault-propagation results should not be interpreted as attributing
the entire logical-$Z$ curve in
Fig.~\ref{fig:833_z_biased_trotter} to a single fault mechanism. That
simulation also includes noisy syndrome extraction, the
Chao--Reichardt recovery procedure, and higher-order fault events.
Rather, the enumeration identifies a structural limitation of direct
Trotter flagging: even when two flag outcomes and a full stabilizer
syndrome are available, some single faults remain logically
indistinguishable or undetected.

These observations indicate that the difficulty is not merely
detecting an error after it has propagated through the entire
high-weight Trotter circuit. Once such propagation has occurred, the
two flag outcomes and the final stabilizer syndrome may no longer
uniquely determine the logical coset of the residual data error.

We therefore pursue two complementary lines of investigation. First,
we test whether damaging faults can be removed through verification
or conservative rejection rather than corrected after the complete
propagation has occurred. Sections~\ref{sec:833_clinr} and
\ref{sec:833_flag_postselection} examine, respectively, a
CliNR-based resource-verification construction~\cite{delfosse2025low} and direct flag
postselection, with the logical-$X$ and logical-$Z$ failure channels
evaluated separately. Second, in
Sec.~\ref{sec:833_asymmetric_noise}, we separate the physical error
scales of the CNOT parity network and the central phase rotation,
compare the direct circuit with and without a noisy
Chao--Reichardt recovery round, and use the single-fault analysis to
construct a diagnostic protected limit. The latter experiment
identifies the restricted set of protected locations under which
both logical sectors recover pseudo-threshold behavior.

\subsection{CliNR-based encoded Pauli rotations}
\label{sec:833_clinr}

The fault-propagation analysis above indicates that detecting an
error only after it has traversed the complete encoded Pauli-rotation
circuit may provide insufficient information for recovery. We
therefore investigate Clifford noise reduction (CliNR), a
gate-teleportation-based protocol that attempts to detect faults in
an auxiliary resource state before that state interacts with the
input data~\cite{delfosse2025low}.
We note that this is the only part of the paper that is necessarily
restricted to Clifford Trotter circuits; in all other places, our 
techniques can be generalized to any angle in the Trotter circuit.

For an \(n\)-qubit Clifford circuit \(C\), CliNR uses three
\(n\)-qubit registers and one additional measurement ancilla. The
second and third registers are initialized as \(n\) Bell pairs, after
which \(C\) is applied to the third register. Stabilizers of the
resulting resource state are then measured before the resource state
is coupled to the input register. If any selected resource-stabilizer
measurement returns the \(-1\) outcome, the resource preparation is
rejected and restarted. Since this verification is performed
offline, a rejected resource state does not corrupt the
input data.

\begin{figure}[h]
    \centering
    \includegraphics[
        width=\columnwidth
    ]{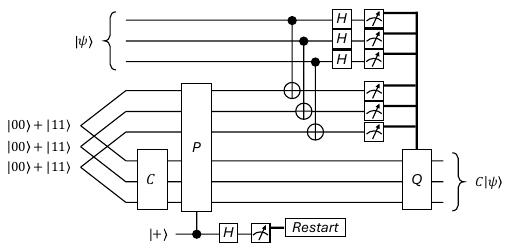}
    \caption{
        CliNR protocol used to implement the Clifford circuit \(C\).
        The circuit is first applied to one half of a Bell-pair
        resource state. The block \(P\) verifies selected stabilizers
        of that resource state before it interacts with the input
        register. A nontrivial verification outcome causes a restart,
        while an accepted resource state is consumed by transversal
        gate teleportation followed by the classically controlled
        Pauli correction \(Q\).
    }
    \label{fig:833_clinr_protocol}
\end{figure}

The accepted resource state is consumed through transversal CNOTs,
single-qubit measurements, and a classically controlled Pauli
correction. If \(o_i\) and \(o_{n+i}\) denote the two measurement
outcomes associated with qubit \(i\), the correction is
\begin{equation}
    Q
    =
    \prod_{i=1}^{n}
    \left(
        C^{\dagger}X_iC
    \right)^{o_{n+i}}
    \left(
        C^{\dagger}Z_iC
    \right)^{o_i}.
    \label{eq:833_clinr_feedforward}
\end{equation}
For Clifford \(C\), every operator appearing in
Eq.~\eqref{eq:833_clinr_feedforward} is a Pauli operator and can be
implemented as a Pauli-frame update.
The CliNR circuit used in this work is summarized in
Fig.~\ref{fig:833_clinr_protocol}. The block labeled \(P\) denotes
the sequence of selected resource-stabilizer measurements. A
nontrivial outcome causes the resource preparation to restart,
whereas an accepted resource state proceeds to gate teleportation
and the correction \(Q\).

Before applying \(C\), the \(n\) Bell pairs possess the \(2n\)
independent stabilizer generators
\begin{equation}
    X_{L,i}X_{R,i},
    \qquad
    Z_{L,i}Z_{R,i},
    \qquad
    i=1,\ldots,n.
    \label{eq:833_bell_resource_stabilizers}
\end{equation}
After applying \(C\) to the right register, these generators become
\begin{equation}
    X_{L,i}
    \left(
        C X_{R,i} C^{\dagger}
    \right),
    \qquad
    Z_{L,i}
    \left(
        C Z_{R,i} C^{\dagger}
    \right).
    \label{eq:833_propagated_resource_stabilizers}
\end{equation}
Only \(r\) of these stabilizers are measured. The original CliNR
construction selects independent resource stabilizers randomly,
whereas its numerical analysis also observes that structured
circuits can require a selection adapted to the circuit shape and
noise profile~\cite{delfosse2025low}. Resource-stabilizer
selection is therefore an important design choice in the present
application.
Our encoded Pauli-rotation circuits act on four data qubits and one
parity ancilla, so the CliNR resource state contains \(n=5\) Bell
pairs and has ten independent stabilizer generators. We use \(r=4\)
resource checks.

For every accepted CliNR shot, we append the same final ideal cleanup
used in the preceding circuit-level simulations. The cleanup consists
of an ideal full syndrome-extraction pass followed by the corresponding
weight-one recovery from \(\mathcal{D}_{\mathrm{LUT}}\). It is applied
after the gate-teleportation circuit and its classically controlled
Pauli correction \(Q\), but before the final propagated-logical-observable
measurement. Unless stated otherwise, all CliNR logical error rates
reported below include this final ideal cleanup. As before, the cleanup
is used only to define the final simulation boundary and is not counted
as part of the noisy CliNR protocol.

All CliNR results reported in this subsection are obtained from
\(N=10^{7}\) attempted Monte Carlo shots for each sampled physical
error rate and each logical ensemble. The logical-\(X\) and
logical-\(Z\) experiments are simulated separately using their
corresponding initial logical states and final propagated logical
observables. For each logical ensemble, the conditional logical
failure rate and acceptance probability are defined as
\begin{align}
    P_L
    &=
    \frac{
        N_{\mathrm{fail,acc}}
    }{
        N_{\mathrm{acc}}
    },
    \label{eq:833_conditional_logical_failure_rate}
    \\
    P_{\mathrm{acc}}
    &=
    \frac{
        N_{\mathrm{acc}}
    }{
        N
    },
    \qquad
    N=10^{7},
    \label{eq:833_acceptance_probability}
\end{align}
where \(N_{\mathrm{acc}}\) is the number of accepted shots and
\(N_{\mathrm{fail,acc}}\) is the number of accepted shots classified
as logical failures.

Among the choices tested, four checks of the Pauli type matching the
rotation axis produced the best performance: \(Z\)-type checks for
the \(Z\)-type rotation and \(X\)-type checks for the \(X\)-type
rotation. Other tested choices resulted in substantially larger
logical error rates. A systematic optimization of the
resource-stabilizer set remains open.

We first consider
\begin{equation}
    \overline{U}_{ZZ}
    =
    \exp\left(
        -\frac{i\pi}{4}
        \overline{Z_2} \, \overline{Z_3}
    \right),
    \qquad
    \overline{Z_2} \, \overline{Z_3}
    =
    Z_2Z_4Z_5Z_6.
    \label{eq:833_clinr_z2z3_rotation}
\end{equation}
This rotation provides a nontrivial two-logical-qubit benchmark. Its
unencoded counterpart,
\begin{equation}
    U_{ZZ}
    =
    \exp\left(
        -\frac{i\pi}{4}Z_2Z_3
    \right),
    \label{eq:833_unencoded_z2z3_rotation}
\end{equation}
requires an entangling implementation,
\begin{equation}
    U_{ZZ}
    =
    \operatorname{CNOT}_{2,3}
    P_3
    \operatorname{CNOT}_{2,3},
    \label{eq:833_unencoded_z2z3_circuit}
\end{equation}
up to an irrelevant global phase. This makes it a more informative
unencoded reference than a single-qubit \(Z\) rotation, which would
reduce to a phase gate alone.

The encoded circuit has the same parity-computation pattern as the
unflagged rotation in Fig.~\ref{fig:833_plain_trotter_circuit}, with
the support \(\{1,5,6,7\}\) replaced by
\(\{2,4,5,6\}\). Both
\(\overline{Z_2}=Z_1Z_5Z_6Z_7\) and
\(\overline{Z_2} \, \overline{Z_3}=Z_2Z_4Z_5Z_6\) have weight four.
Consequently, their direct encoded implementations have the same
number and pattern of parity-computation gates and exhibit similar
circuit-level behavior.

\begin{figure*}[t]
    \centering
    \begin{minipage}{0.49\textwidth}
        \centering
        \includegraphics[
            width=\linewidth
        ]{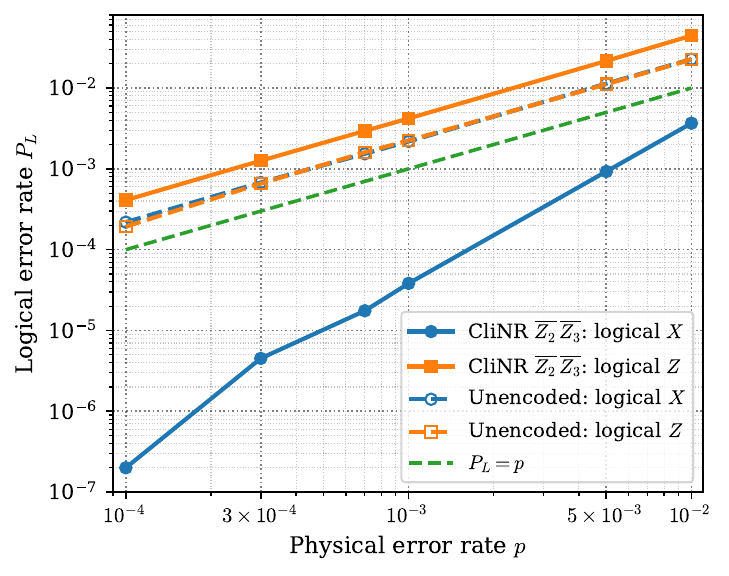}

        \smallskip
        \textbf{(a)}
    \end{minipage}
    \hfill
    \begin{minipage}{0.49\textwidth}
        \centering
        \includegraphics[
            width=\linewidth
        ]{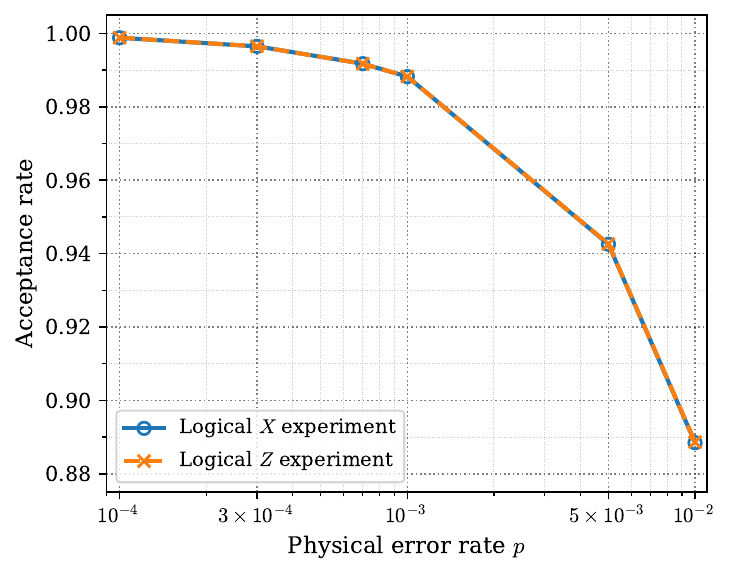}

        \smallskip
        \textbf{(b)}
    \end{minipage}
    \caption{
        CliNR implementation of
        \(\exp(-i\pi\overline{Z_2} \, \overline{Z_3}/4)\) using
        \(r=4\) \(Z\)-type resource-stabilizer checks.
        \textbf{(a)} Logical-\(X\) and logical-\(Z\) failure rates,
        compared with the corresponding unencoded two-qubit Pauli
        rotation and the reference \(P_L=p\).
        \textbf{(b)} Acceptance probabilities for the logical-$X$ and
         logical-$Z$ experiments.
    }
    \label{fig:833_clinr_z2z3}
\end{figure*}

Figure~\ref{fig:833_clinr_z2z3} shows the logical error rates and
acceptance probability for the CliNR implementation of
\(\overline{U}_{ZZ}\). The logical-\(X\) failure rate is reduced by
at least approximately one order of magnitude relative to the
unencoded two-qubit rotation throughout the sampled range, with a
larger separation at the lowest simulated error rates. In contrast,
the logical-\(Z\) failure rate remains above the unencoded reference
and is comparable in scale to that of the direct encoded Trotter
implementation. CliNR therefore produces strongly asymmetric
suppression rather than simultaneous suppression of both logical
channels.

The acceptance probability decreases smoothly with $p$, from nearly
unity at $p=10^{-4}$ to approximately $0.89$ at $p=10^{-2}$. The
acceptance probabilities obtained from the logical-$X$ and
logical-$Z$ experiments differ only slightly and therefore appear
overlapped at the resolution of the right panel of Fig.~\ref{fig:833_clinr_z2z3}.
This agreement is expected because the same CliNR resource-state
preparation and verification circuit is used in both experiments;
the two ensembles differ primarily in their initial logical states
and final logical measurements. Thus, the reduction in the
logical-$X$ failure rate does not arise from a substantially more
selective acceptance condition for that ensemble. Nevertheless, the
rejection cost must be included when comparing the total sampling
overhead with direct implementations.

The observed asymmetry can be understood from the propagated
resource stabilizers. Let
\begin{equation}
    P_Z
    =
    \prod_{j\in\mathcal{S}_{ZZ}} Z_j,
    \qquad
    \mathcal{S}_{ZZ}
    =
    \{2,4,5,6\},
    \label{eq:833_clinr_z_generator}
\end{equation}
so that \(C=\exp(-i\pi P_Z/4)\). Ignoring an overall Pauli phase,
conjugation gives
\begin{align}
    C^{\dagger}Z_iC
    &=
    Z_i,
    \label{eq:833_clinr_z_conjugation_z}
    \\
    C^{\dagger}X_iC
    &\doteq
    Y_i
    \prod_{\substack{
        j\in\mathcal{S}_{ZZ}\\
        j\neq i
    }}
    Z_j,
    \label{eq:833_clinr_z_conjugation_x}
    \\
    C^{\dagger}Y_iC
    &\doteq
    X_i
    \prod_{\substack{
        j\in\mathcal{S}_{ZZ}\\
        j\neq i
    }}
    Z_j,
    \label{eq:833_clinr_z_conjugation_y}
\end{align}
where \(\doteq\) denotes equality up to an overall phase. The
\(Z\)-type resource checks commute with residual \(Z\) faults but
detect many faults containing an \(X\) or \(Y\) component. The
accepted ensemble is therefore biased toward residual \(Z\)-type
errors.

Faults in the transversal teleportation layer can also flip one or
both classical measurement outcomes. Through
Eq.~\eqref{eq:833_clinr_feedforward}, such flips insert one of the
propagated corrections \(C^{\dagger}Z_iC\),
\(C^{\dagger}X_iC\), or \(C^{\dagger}Y_iC\). In particular, a
measurement-record error can produce a correlated Pauli of the form
\begin{equation}
    E_i
    \doteq
    Y_i
    \prod_{\substack{
        j\in\mathcal{S}_{ZZ}\\
        j\neq i
    }}
    Z_j.
    \label{eq:833_clinr_correlated_yz_error}
\end{equation}
This operator has the same syndrome as a weight-one \(X_i\) error.
The final weight-one recovery therefore gives
\begin{equation}
    X_iE_i
    \doteq
    \prod_{j\in\mathcal{S}_{ZZ}}Z_j
    =
    \overline{Z_2} \, \overline{Z_3}.
    \label{eq:833_clinr_cleanup_to_logical_z}
\end{equation}
Thus, the ideal cleanup can convert this residual correlated error
into a pure logical-\(Z\)-type representative. It removes the
correctable component and makes the remaining logical coset explicit.
This mechanism provides one important contribution to the strong
logical-sector asymmetry observed after final cleanup: the
correctable syndrome component is removed, while a nontrivial
logical-\(Z\) component can remain.

To assess the contribution of this boundary recovery, we also performed
an otherwise identical auxiliary simulation with the final ideal
cleanup removed. Without cleanup, the logical-$X$ curve does not
exhibit a pseudo-threshold. This comparison, not shown, demonstrates
that the suppression observed in Fig.~\ref{fig:833_clinr_z2z3} results
from the combination of CliNR resource verification and the final
error-correction step, rather than from CliNR verification alone. The
ideal cleanup remains a simulation-boundary operation and should not
be interpreted as a hardware-realistic component of the noisy CliNR
protocol.

The strong dependence on the resource checks also shows that their
selection is not freely adjustable. For a \(Z\)-type rotation,
the generators
\begin{equation}
    Z_{L,i}
    \left(
        C Z_{R,i}C^{\dagger}
    \right)
    =
    Z_{L,i}Z_{R,i}
    \label{eq:833_clinr_simple_z_resource_checks}
\end{equation}
remain weight-two \(Z\)-type checks. By contrast, the conjugated
\(X\)-type generators become higher-weight mixed-Pauli operators.
Choosing only the simple \(Z\)-type generators efficiently rejects
one class of faults but leaves the complementary \(Z\)-type channel
weakly constrained. No tested set of four resource checks
simultaneously suppressed both logical sectors.

\begin{figure}[h]
    \centering
    \includegraphics[
        width=\columnwidth
    ]{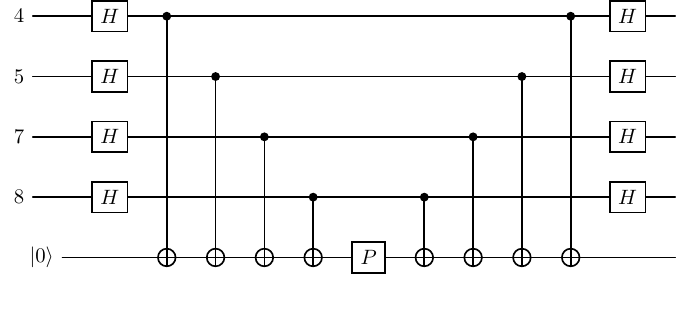}
    \caption{
        Direct encoded implementation of
        \(\exp(-i\pi\overline{X_1}/4)\), where
        \(\overline{X_1}=X_4X_5X_7X_8\). Hadamard gates rotate the
        four-qubit \(X\)-type support into the \(Z\) basis before the
        parity-computation circuit and rotate it back afterward.
    }
    \label{fig:833_x1_trotter_circuit}
\end{figure}

To test whether this asymmetry follows the rotation axis rather than
a special property of
\(\overline{Z_2} \, \overline{Z_3}\), we next consider
\begin{equation}
    \overline{U}_{X}
    =
    \exp\left(
        -\frac{i\pi}{4}\overline{X_1}
    \right),
    \qquad
    \overline{X_1}
    =
    X_4X_5X_7X_8.
    \label{eq:833_clinr_x1_rotation}
\end{equation}
The corresponding direct encoded parity circuit is shown in
Fig.~\ref{fig:833_x1_trotter_circuit}. Hadamard gates map the
\(X\)-type Pauli support to and from the \(Z\) basis around the
parity-computation network.

\begin{figure*}[t]
    \centering
    \begin{minipage}{0.49\textwidth}
        \centering
        \includegraphics[
            width=\linewidth
        ]{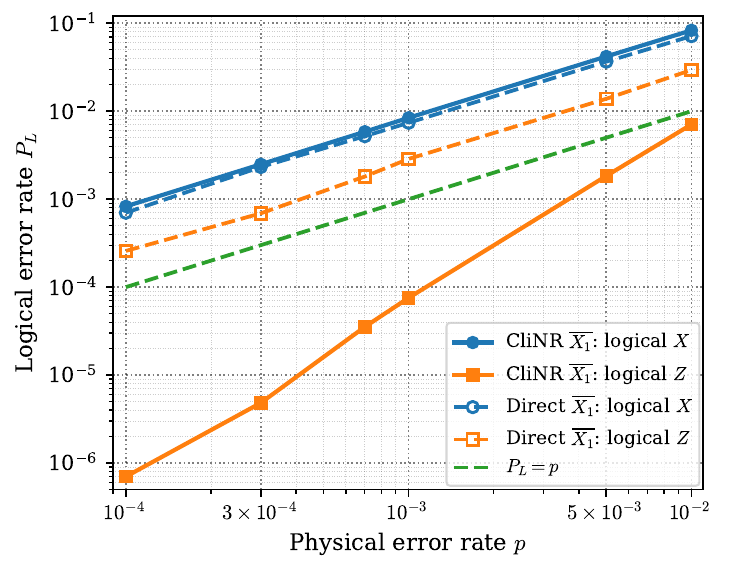}

        \smallskip
        \textbf{(a)}
    \end{minipage}
    \hfill
    \begin{minipage}{0.49\textwidth}
        \centering
        \includegraphics[
            width=\linewidth
        ]{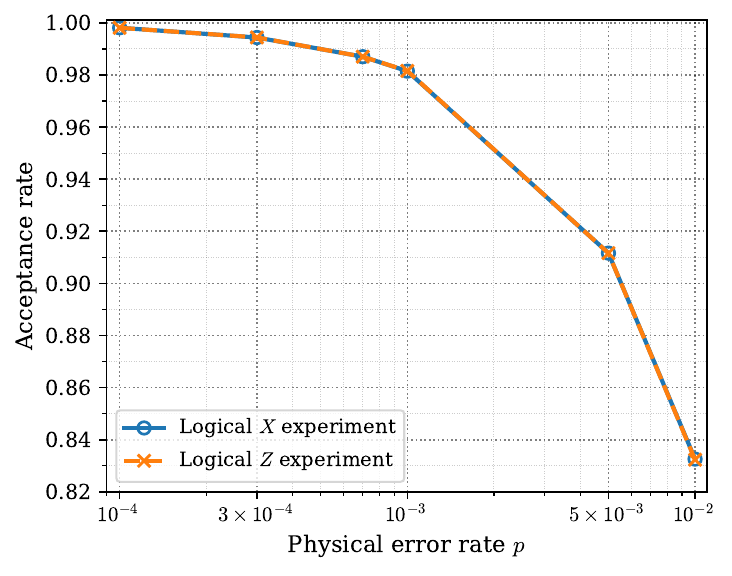}

        \smallskip
        \textbf{(b)}
    \end{minipage}
    \caption{
        CliNR implementation of
        \(\exp(-i\pi\overline{X_1}/4)\) using \(r=4\)
        \(X\)-type resource-stabilizer checks.
        \textbf{(a)} Logical-\(X\) and logical-\(Z\) failure rates,
        compared with the direct encoded implementation with final
        ideal cleanup and with \(P_L=p\).
        \textbf{(b)} Acceptance probabilities for the logical-$X$ and
         logical-$Z$ experiments. 
    }
    \label{fig:833_clinr_x1}
\end{figure*}

For this circuit, we again use \(r=4\), but select four \(X\)-type
resource-stabilizer checks. Figure~\ref{fig:833_clinr_x1} shows a
reversal of the preceding behavior. The logical-\(Z\) failure rate
is reduced by multiple orders of magnitude relative to the direct
encoded implementation. The logical-\(X\) failure rate, however,
remains approximately equal to the direct encoded result over the
sampled range and does not cross below \(P_L=p\).

The direct encoded circuit is used as the baseline in
Fig.~\ref{fig:833_clinr_x1}, rather than the unencoded rotation. The
unencoded implementation is simply
\begin{equation}
    \exp\left(
        -\frac{i\pi}{4}X_1
    \right)
    =
    H_1P_1H_1,
    \label{eq:833_unencoded_x_rotation}
\end{equation}
up to a global phase, and contains no two-qubit gates. It is
therefore too weak a reference for isolating the effect of CliNR on
the encoded high-weight interaction. The direct encoded baseline
uses the same final ideal-cleanup convention as the CliNR
simulation.

The \(X\)-axis result follows the conjugate version of the preceding
mechanism. With
\begin{equation}
    P_X
    =
    \prod_{j\in\mathcal{S}_{X}}X_j,
    \qquad
    \mathcal{S}_{X}
    =
    \{4,5,7,8\},
    \label{eq:833_clinr_x_generator}
\end{equation}
and \(C=\exp(-i\pi P_X/4)\), one has, up to phase,
\begin{align}
    C^{\dagger}X_iC
    &=
    X_i,
    \label{eq:833_clinr_x_conjugation_x}
    \\
    C^{\dagger}Z_iC
    &\doteq
    Y_i
    \prod_{\substack{
        j\in\mathcal{S}_{X}\\
        j\neq i
    }}
    X_j,
    \label{eq:833_clinr_x_conjugation_z}
    \\
    C^{\dagger}Y_iC
    &\doteq
    Z_i
    \prod_{\substack{
        j\in\mathcal{S}_{X}\\
        j\neq i
    }}
    X_j.
    \label{eq:833_clinr_x_conjugation_y}
\end{align}
Consequently, the \(X\)-type resource checks preferentially reject
faults with \(Z\) or \(Y\) components while allowing more residual
\(X\)-type faults to survive. Final weight-one recovery can map a
correlated operator of the form
\begin{equation}
    Y_i
    \prod_{\substack{
        j\in\mathcal{S}_{X}\\
        j\neq i
    }}
    X_j
    \label{eq:833_clinr_correlated_yx_error}
\end{equation}
into the logical-\(X\)-type representative
\begin{equation}
    Z_i
    \left(
        Y_i
        \prod_{\substack{
            j\in\mathcal{S}_{X}\\
            j\neq i
        }}
        X_j
    \right)
    \doteq
    \prod_{j\in\mathcal{S}_{X}}X_j
    =
    \overline{X_1}.
    \label{eq:833_clinr_cleanup_to_logical_x}
\end{equation}
This is the \(X\)-type counterpart of
Eq.~\eqref{eq:833_clinr_cleanup_to_logical_z} and explains the
reversal of the two logical failure channels.

The acceptance probability for the $\overline{X_1}$ experiment
decreases from nearly unity at $p=10^{-4}$ to approximately $0.83$
at $p=10^{-2}$. As in the right panel of Fig.~\ref{fig:833_clinr_z2z3}, the logical-$X$ and logical-$Z$ acceptance probabilities differ only slightly and consequently appear overlapped in the right panel of Fig.~\ref{fig:833_clinr_x1}. This shows that the strong reversal between the two logical-error channels is not caused by a substantially different postselection rate for the two input ensembles. The somewhat lower high-noise acceptance rate than in the $\overline{Z_2} \, \overline{Z_3}$ experiment is consistent with the additional Hadamard locations in the $X$-type implementation.

Taken together, the two CliNR experiments demonstrate substantial
noise reduction in one logical sector at a time, but not
simultaneous suppression of both sectors. Matching the resource
checks to the rotation axis strongly suppresses faults that
anticommute with those checks, while faults aligned with the same
axis remain comparatively difficult to detect. Moreover, noisy
teleportation measurements can alter the feedforward operation
\(Q\) and generate correlated output Paulis. For the present
\(\llbr 8,3,3\rrbr\) implementation, the freedom to choose resource
stabilizers is therefore constrained by the conjugation action of
\(C\), and no tested \(r=4\) selection eliminates both complementary
logical channels. This motivates the flagged-postselection strategy
examined next.

\subsection{Flagged postselection of encoded Pauli rotations}
\label{sec:833_flag_postselection}

The preceding results show that both direct flag-conditioned recovery
and CliNR resource verification can be limited by ambiguity in the
residual error record. We therefore consider a simpler
postselection strategy. Instead of attempting to infer and correct
the propagated data error associated with a triggered flag, we reject
the shot whenever either Trotter flag is raised. If \(f_L\) and
\(f_R\) denote the outcomes of the two flag measurements, a shot is
accepted if and only if
\begin{equation}
    f_L=f_R=0.
    \label{eq:833_flag_postselection_condition}
\end{equation}
Thus, all singly and doubly flagged events are rejected, regardless
of whether the flag was caused by a damaging propagated error, a
correctable weight-one error, or a benign fault that acts trivially
on the data. This deliberately conservative rule avoids the
flag-conditioned decoding ambiguities identified above, at the cost
of rejecting false alarms and otherwise correctable events.

The flag-postselection experiments use the same sampling convention
as the CliNR simulations. For each sampled physical error rate and
each logical ensemble, we use \(N=10^{7}\) attempted Monte Carlo
shots. The reported logical failure rates and acceptance
probabilities are evaluated according to
Eqs.~\eqref{eq:833_conditional_logical_failure_rate}
and~\eqref{eq:833_acceptance_probability}, respectively.

For every accepted shot, the flagged Trotter circuit is followed by
the same final ideal cleanup used in the preceding simulations. The
cleanup consists of ideal full-syndrome extraction and the
corresponding weight-one recovery from
\(\mathcal{D}_{\mathrm{LUT}}\), followed by measurement of the
ideally propagated logical observables, i.e.,
\begin{equation}
    \begin{aligned}
        \text{flagged rotation}
        &\longrightarrow
        \text{postselection}
        \\
        &\longrightarrow
        \text{final ideal cleanup}.
    \end{aligned}
    \label{eq:833_flag_postselection_architecture}
\end{equation}
The flag-postselection layer itself is compatible with a
near-term detection strategy because it requires only classical
discarding of flagged shots. The ideal cleanup, however, remains a
simulation-boundary operation and is not treated as part of a
hardware-realistic noisy protocol.

We also examined an alternative \(X\)-type flag construction designed
to detect \(X\) faults on the parity ancilla during the first half of
the Trotter circuit. Such faults are particularly relevant because an
\(X\) component present before the phase gate is transformed into a
\(Y\) component and can subsequently propagate through the remaining
parity network. Detecting this propagation requires additional flag
interactions, so the \(X\)-type construction introduces more
two-qubit gates and hence more faulty circuit locations than the
\(Z\)-type construction.

In auxiliary simulations not shown here, the additional \(X\)-type
flag circuit did not provide a corresponding reduction in the logical
failure rates. Among the flag layouts tested, the two-\(Z\)-flag
construction produced the best overall performance. We therefore use
the same two \(Z\)-type flags for both encoded rotations studied below. This choice is based on the observed circuit-level performance
rather than a random selection of the flag type. The comparison also
illustrates a general tradeoff in direct Trotter flagging: increasing
the set of detectable propagated faults requires additional
entangling gates, which themselves create new fault locations. For the phase gate \(P\), an \(X\) component present before the phase gate is transformed into a \(Y\) component.

We first apply this construction to
\begin{equation}
    \overline{U}_{ZZ}
    =
    \exp\left(
        -\frac{i\pi}{4}
        \overline{Z_2} \, \overline{Z_3}
    \right),
    \qquad
    \overline{Z_2} \, \overline{Z_3}
    =
    Z_2Z_4Z_5Z_6.
    \label{eq:833_flag_postselection_z2z3_rotation}
\end{equation}
The flagged circuit has the same structure as the flagged
weight-four \(Z\)-type rotation in
Fig.~\ref{fig:833_flagged_trotter_circuit}, with the data support
\(\{1,5,6,7\}\) replaced by \(\{2,4,5,6\}\). The two \(Z\)-type
flags monitor the parity-computation and parity-uncomputation
portions of the circuit. Unlike the flag-conditioned procedure used
in the biased-noise experiment, no lookup table is applied when a
flag is raised; the corresponding shot is discarded immediately.

\begin{figure*}[t]
    \centering
    \begin{minipage}{0.49\textwidth}
        \centering
        \includegraphics[
            width=\linewidth
        ]{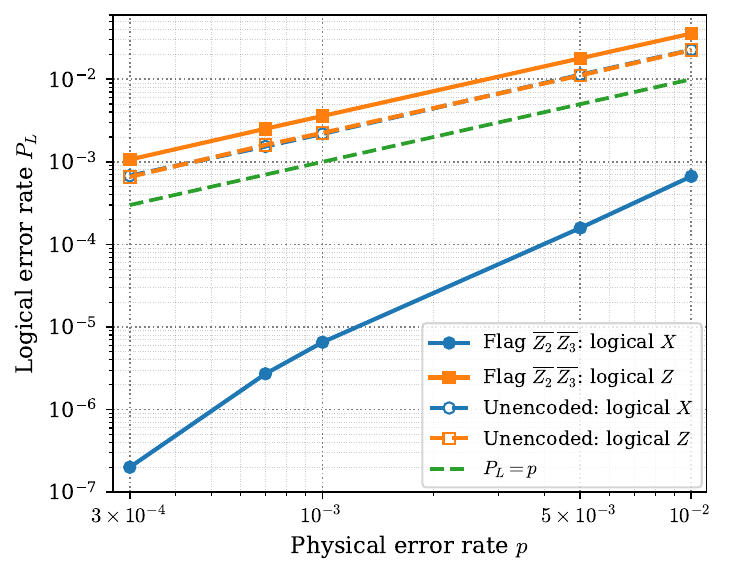}

        \smallskip
        \textbf{(a)}
    \end{minipage}
    \hfill
    \begin{minipage}{0.49\textwidth}
        \centering
        \includegraphics[
            width=\linewidth
        ]{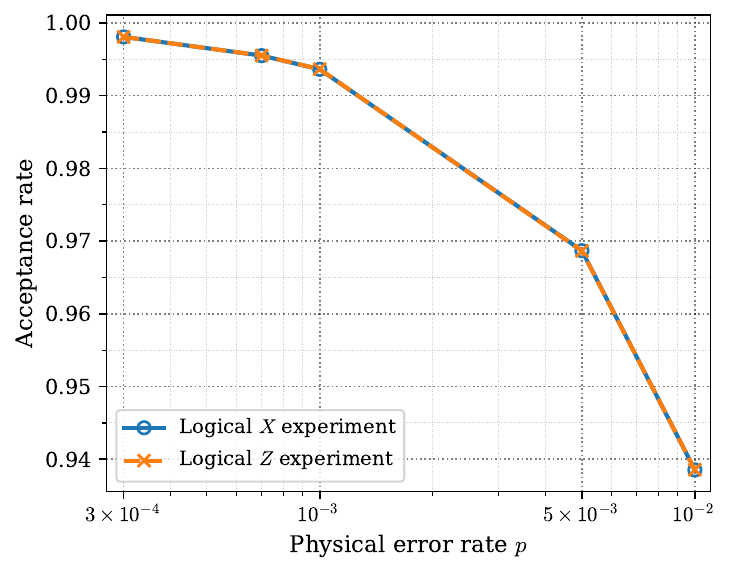}

        \smallskip
        \textbf{(b)}
    \end{minipage}
    \caption{
        Flag-postselected implementation of
        \(\exp(-i\pi\overline{Z_2} \, \overline{Z_3}/4)\), followed by
        final ideal cleanup.
        \textbf{(a)} Logical-\(X\) and logical-\(Z\) failure rates,
        compared with the corresponding unencoded two-qubit Pauli
        rotation and the reference \(P_L=p\).
        \textbf{(b)} Acceptance probabilities for the logical-\(X\)
        and logical-\(Z\) experiments. The plotted range begins at \(p=3\times10^{-4}\) because one or more lower-noise points yielded zero observed logical-\(X\) failures. These zero-count estimates are omitted rather than plotted on the logarithmic scale.
    }
    \label{fig:833_flag_postselection_z2z3}
\end{figure*}

The logical error rates and acceptance probabilities are shown in
Fig.~\ref{fig:833_flag_postselection_z2z3}. Flag postselection
suppresses the logical-\(X\) failure rate particularly strongly. Over
the common parameter range, it is approximately another order of
magnitude below the corresponding CliNR result in
Fig.~\ref{fig:833_clinr_z2z3}, and it remains well below both the
unencoded logical-\(X\) reference and \(P_L=p\). In contrast, the
logical-\(Z\) failure rate remains above the unencoded reference.
Consequently, the additional logical-\(X\) suppression does not
resolve the persistent logical-\(Z\) channel.

The acceptance probability remains close to unity throughout most
of the sampled range and decreases to approximately \(0.94\) at
\(p=10^{-2}\). The logical-\(X\) and logical-\(Z\) experiments use
the same flagged Trotter circuit and acceptance condition. Their
acceptance probabilities therefore differ only slightly and appear
overlapped in the right panel of
Fig.~\ref{fig:833_flag_postselection_z2z3}. In particular, the
strong difference between the two conditional logical error rates
does not arise from applying a substantially more selective
postselection rule to one logical ensemble.

The flag-postselection acceptance rate is also higher than the
corresponding CliNR acceptance rate in
Fig.~\ref{fig:833_clinr_z2z3}. This is consistent with the smaller
verification circuit: the flag-postselection construction rejects
only faults that alter one of the two flag outcomes, whereas CliNR
measures four resource stabilizers and restarts if any one of them
returns a nontrivial value. The higher acceptance rate does not,
however, imply that every accepted event is harmless.

To examine the mechanism, we enumerate single Pauli faults at every
faultable location of the flagged parity circuit and separate the
resulting errors according to whether at least one flag is raised.
The flagged subset contains many damaging correlated errors,
including faults whose data component is the complete weight-four
operator
\begin{equation}
    P_Z
    =
    Z_2Z_4Z_5Z_6
    =
    \overline{Z_2} \, \overline{Z_3}.
    \label{eq:833_flag_postselection_z_logical_operator}
\end{equation}
Postselection therefore removes a substantial collection of
first-order logical and correlated faults. It also rejects benign
events, including faults whose final data component is the identity
or a correctable error. These events contribute to the rejection
overhead but not directly to the conditional logical error rate.

The unflagged subset nevertheless still contains occurrences of
\(P_Z\), as well as correlated errors of the forms
\begin{align}
    E^{(Y)}_i
    &\doteq
    Y_i
    \prod_{\substack{
        j\in\mathcal{S}_{ZZ}\\
        j\neq i
    }}
    Z_j,
    \label{eq:833_unflagged_yz_pattern}
    \\
    E^{(X)}_i
    &\doteq
    X_i
    \prod_{\substack{
        j\in\mathcal{S}_{ZZ}\\
        j\neq i
    }}
    Z_j,
    \label{eq:833_unflagged_xz_pattern}
\end{align}
where \(\mathcal{S}_{ZZ}=\{2,4,5,6\}\) and \(\doteq\) denotes
equality up to an overall phase. The fault enumeration also produces
many unflagged weight-one data errors. These weight-one errors are
removed by the final ideal cleanup and therefore do not appear as
logical failures in the reported curves.

More subtly, some correlated accepted errors are mapped by the
weight-one recovery into a logical-\(Z\)-type representative. For
example, \(E^{(Y)}_i\) has the same syndrome as a weight-one \(X_i\)
error, and hence
\begin{equation}
    X_iE^{(Y)}_i
    \doteq
    \prod_{j\in\mathcal{S}_{ZZ}}Z_j
    =
    \overline{Z_2} \, \overline{Z_3}.
    \label{eq:833_flag_cleanup_to_logical_z}
\end{equation}
The cleanup removes the correctable syndrome component but leaves
the nontrivial logical coset. This mechanism provides one important contribution to the strong logical-sector asymmetry observed after final cleanup. It shows how the weight-one recovery can remove a correctable syndrome component while leaving the accepted error in a nontrivial logical-\(Z\) coset.

Although CliNR and direct flag postselection use different rejection
mechanisms, their accepted error ensembles therefore exhibit a
similar qualitative structure. CliNR rejects faults that
anticommute with the selected resource stabilizers, whereas direct
postselection rejects faults that change a flag outcome. In both
cases, many transverse \(X/Y\)-containing errors are filtered out,
while some axis-aligned logical operators and correlated errors
remain accepted. Final ideal cleanup removes the surviving
weight-one component, revealing a residual logical channel aligned
with the rotation axis.

\begin{figure*}[t]
    \centering
    \includegraphics[
        width=0.82\textwidth
    ]{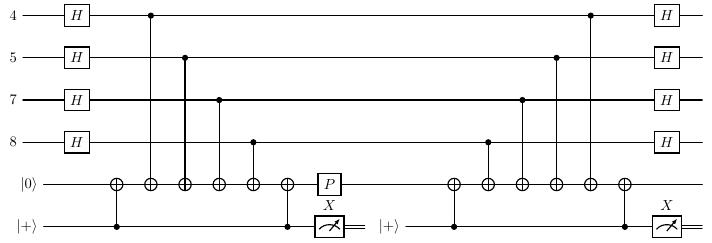}
    \caption{
    Flagged implementation of
    \(\exp(-i\pi\overline{X_1}/4)\), where
    \(\overline{X_1}=X_4X_5X_7X_8\). Hadamard gates rotate the
    \(X\)-type support into and out of the \(Z\) basis. The two
    \(Z\)-type flag ancillas monitor the parity-computation and
    parity-uncomputation portions of the circuit. In the
    postselection experiment, the shot is rejected whenever either
    flag is raised.
    }
    \label{fig:833_flag_postselection_x1_circuit}
\end{figure*}

We next apply the same postselection principle to
\begin{equation}
    \overline{U_X}
    =
    \exp\left(
        -\frac{i\pi}{4}\overline{X_1}
    \right),
    \qquad
    \overline{X_1}
    =
    X_4X_5X_7X_8.
    \label{eq:833_flag_postselection_x1_rotation}
\end{equation}
The flagged circuit is shown in
Fig.~\ref{fig:833_flag_postselection_x1_circuit}. As in the direct
encoded implementation, Hadamard gates rotate the four-qubit
\(X\)-type support into and out of the \(Z\) basis. Two flag
ancillas monitor the parity network, and the shot is accepted only
when neither flag is raised.

\begin{figure*}[t!]
    \centering
    \begin{minipage}{0.49\textwidth}
        \centering
        \includegraphics[
            width=\linewidth
        ]{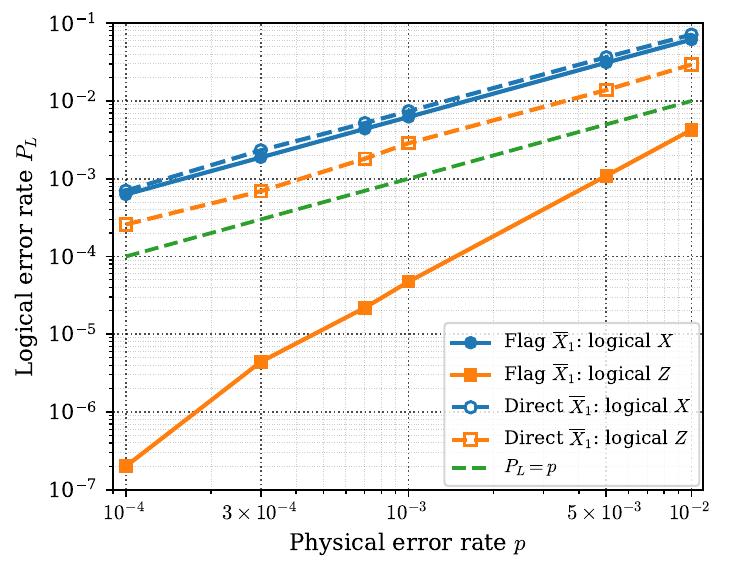}

        \smallskip
        \textbf{(a)}
    \end{minipage}
    \hfill
    \begin{minipage}{0.49\textwidth}
        \centering
        \includegraphics[
            width=\linewidth
        ]{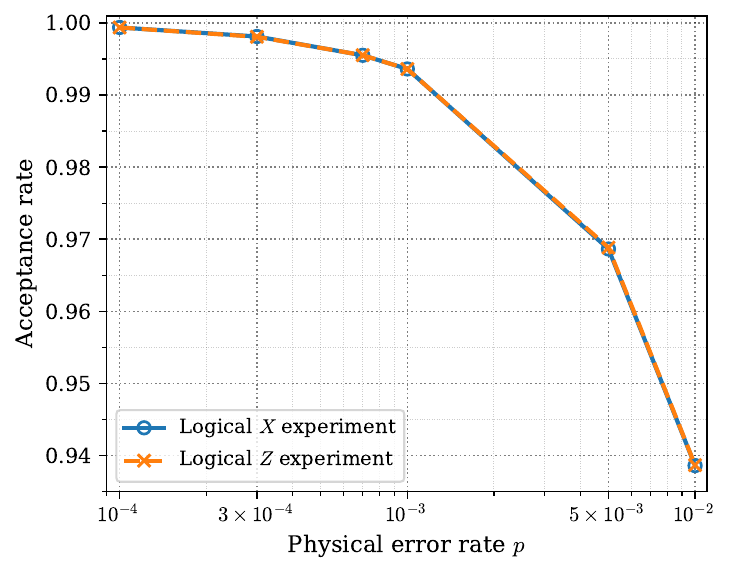}

        \smallskip
        \textbf{(b)}
    \end{minipage}
    \caption{
        Flag-postselected implementation of
        \(\exp(-i\pi\overline{X_1}/4)\), followed by final ideal
        cleanup.
        \textbf{(a)} Logical-\(X\) and logical-\(Z\) failure rates,
        compared with the direct encoded implementation using the
        same final-cleanup convention and with \(P_L=p\).
        \textbf{(b)} Acceptance probabilities for the logical-\(X\)
        and logical-\(Z\) experiments. Their difference is very small,
        so the two curves appear overlapped.
    }
    \label{fig:833_flag_postselection_x1}
\end{figure*}

Figure~\ref{fig:833_flag_postselection_x1} shows the corresponding
logical error rates and acceptance probabilities. The logical-sector
asymmetry is reversed relative to the
\(\overline{Z_2} \, \overline{Z_3}\) experiment. The logical-\(Z\)
failure rate is reduced by multiple orders of magnitude relative to
the direct encoded implementation, whereas the logical-\(X\)
failure rate remains on approximately the same scale as the direct
encoded result and remains above \(P_L=p\).

Compared with the CliNR implementation of
\(\overline{X_1}\) in Fig.~\ref{fig:833_clinr_x1}, flag
postselection provides only a modest further reduction in the
logical-\(Z\) failure rate. This differs from the
\(\overline{Z_2} \, \overline{Z_3}\) case, in which direct
postselection reduces the logical-\(X\) failure rate by approximately
an additional order of magnitude relative to CliNR.

The acceptance probability again remains close to unity at low
physical error rates and is approximately \(0.94\) at
\(p=10^{-2}\). The logical-\(X\) and logical-\(Z\) acceptance
probabilities differ only slightly and appear overlapped in the right
panel of Fig.~\ref{fig:833_flag_postselection_x1}. As in the
\(Z\)-type experiment, the contrasting logical-error curves cannot
be attributed to different acceptance conditions for the two
ensembles.

The reversed behavior is consistent with exchanging the roles of
\(X\) and \(Z\). For the \(X\)-type generator
\begin{equation}
    P_X
    =
    X_4X_5X_7X_8
    =
    \overline{X_1},
    \label{eq:833_flag_postselection_x_logical_operator}
\end{equation}
the flag circuit and final cleanup preferentially remove many errors
with a component transverse to the \(X\) axis, while some accepted
faults remain in, or are mapped into, the logical-\(X\) coset. The
result is strong suppression of logical-\(Z\) failures but only
limited improvement in the logical-\(X\) channel.

Taken together, the two postselection experiments reinforce the
logical-sector asymmetry observed with CliNR. Conservative flag
postselection removes many damaging first-order faults, achieves
higher acceptance probabilities than the corresponding CliNR
implementations, and can substantially suppress the logical channel
transverse to the rotation axis. It does not, however, simultaneously
suppress both logical sectors. Faults aligned with the high-weight
rotation axis can remain unflagged, while final weight-one recovery
can expose their nontrivial logical component. The results therefore
show that neither resource-state verification nor direct Trotter
postselection, in their present forms, provides complete
circuit-level protection for the encoded Pauli rotations considered
here.

\subsection{Asymmetric gate-noise benchmarks and a diagnostic
protected limit}
\label{sec:833_asymmetric_noise}

The preceding results show that the Clifford parity network and the
central phase rotation contribute differently to the two logical
failure channels. We therefore separate their physical error scales.
The CNOT error probability is fixed at
$p_{\mathrm{CNOT}}=10^{-5}$, while the single-qubit depolarizing
error probability following the phase gate,
$p_{\mathrm{phase}}$, is varied independently.
The low CNOT error rate could be justified through a teleported gate
implementation using highly distilled Bell pairs or through an inner
concatenation with a small code that has a transversal CNOT gate.
The final ideal cleanup is retained so that the comparison uses the same terminal
boundary convention as the preceding circuit-level experiments.

Unless stated otherwise, every data point in this subsection is
estimated from $N=10^{7}$ attempted Monte Carlo shots. For an
experiment containing rejection, the reported logical failure
probability is conditioned on acceptance,
\begin{equation}
    \widehat{P}_{L}
    =
    \frac{N_{\mathrm{fail,acc}}}{N_{\mathrm{acc}}}.
    \label{eq:833_asymmetric_conditional_failure_rate}
\end{equation}
where \(N_{\mathrm{acc}}\) is the number of accepted shots and
\(N_{\mathrm{fail,acc}}\) is the number of logical failures among
those shots.

\begin{figure}[t]
    \centering
    \includegraphics[
        width=\columnwidth
    ]{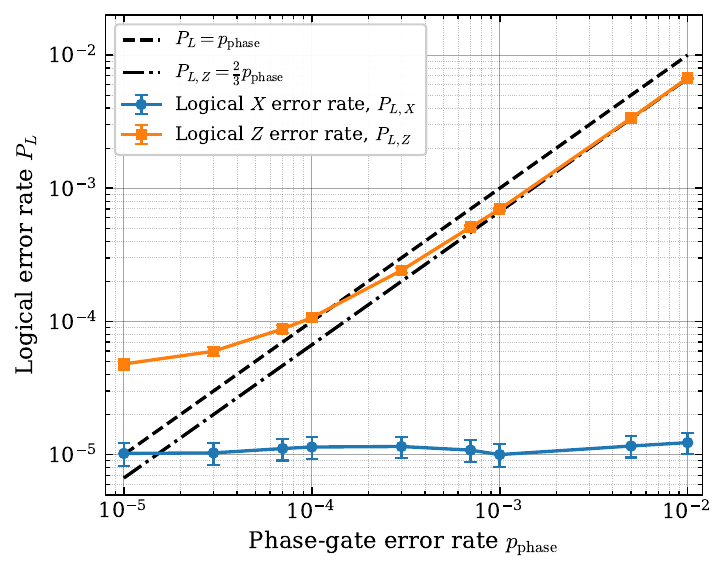}
    \caption{
        Logical failure rates for the encoded Trotter circuit with
        $p_{\mathrm{CNOT}}=10^{-5}$ and independently varied
        $p_{\mathrm{phase}}$, followed by final ideal cleanup.
        At large $p_{\mathrm{phase}}$, the logical-$Z$ failure rate
        approaches approximately
        $2p_{\mathrm{phase}}/3$. At small
        $p_{\mathrm{phase}}$, both the logical-$X$ channel and the
        residual logical-$Z$ contribution are limited by the fixed
        CNOT error scale. Each point is estimated from $N=10^{7}$
        attempted Monte Carlo shots; only point estimates are shown.
    }
    \label{fig:833_asymmetric_direct_trotter}
\end{figure}

We first consider the encoded Trotter circuit without a noisy
error-correction round between the Trotter operation and final
cleanup. The results are shown in
Fig.~\ref{fig:833_asymmetric_direct_trotter}. The logical-$X$ failure
rate remains approximately saturated at the scale set by
$p_{\mathrm{CNOT}}$. At sufficiently small
$p_{\mathrm{phase}}$, the logical-$Z$ failure rate approaches a
similar CNOT-induced floor.

At larger $p_{\mathrm{phase}}$, the logical-$Z$ failure rate is
approximately
\begin{equation}
    P_{L,Z}
    \simeq
    \frac{2}{3}p_{\mathrm{phase}}
    +
    O(p_{\mathrm{CNOT}}).
    \label{eq:833_asymmetric_phase_scaling}
\end{equation}
This coefficient is consistent with the single-qubit depolarizing
channel following the phase gate. Of its three nonidentity Pauli
components, the $Y$ and $Z$ components contain a phase error on the
parity ancilla and can be propagated by the right CNOT ladder into
the logical-$Z$ sector. The remaining contribution at low
$p_{\mathrm{phase}}$ is set by the fixed noisy CNOT locations.

\begin{figure}[t]
    \centering
    \includegraphics[
        width=\columnwidth
    ]{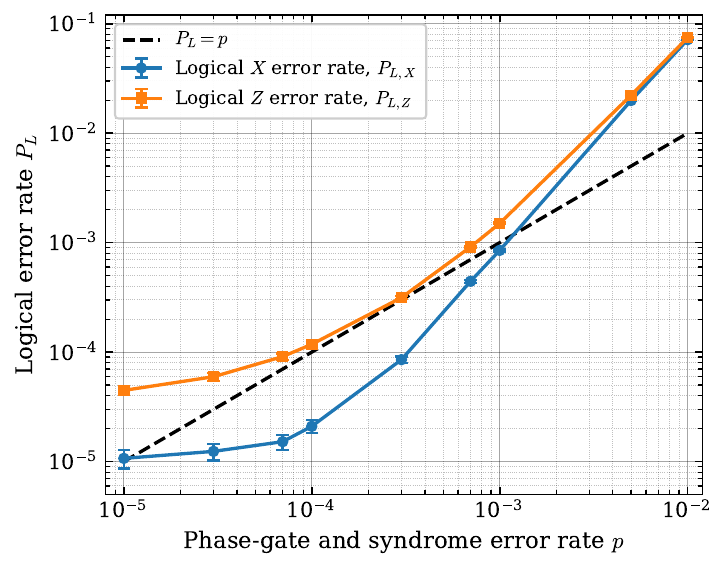}
    \caption{
        Asymmetric encoded-Trotter experiment with
        $p_{\mathrm{CNOT}}=10^{-5}$, followed by one noisy
        Chao--Reichardt error-correction round and final ideal
        cleanup. The additional syndrome-extraction circuitry raises
        the logical failure rates relative to
        Fig.~\ref{fig:833_asymmetric_direct_trotter}. All noisy
        locations in the recovery use
        $p_{\mathrm{QEC}}=p_{\mathrm{phase}}$. Each point is
        estimated from $N=10^{7}$ attempted Monte Carlo shots; only
        point estimates are shown.
    }
    \label{fig:833_asymmetric_trotter_rui}
\end{figure}

We next insert one noisy Chao--Reichardt error-correction round after
the encoded Trotter circuit and before the same final ideal cleanup.
For this experiment, all noisy one- and two-qubit locations in the
Chao--Reichardt recovery use
$p_{\mathrm{QEC}}=p_{\mathrm{phase}}$, while the CNOTs within the
encoded Trotter block remain fixed at
$p_{\mathrm{CNOT}}=10^{-5}$.
The logical failure rates in
Fig.~\ref{fig:833_asymmetric_trotter_rui} are larger than those of
Fig.~\ref{fig:833_asymmetric_direct_trotter}. This degradation is
expected because the syndrome-extraction and recovery circuits add
noisy circuit locations at the variable QEC error scale. Reducing the
error rate of the Trotter CNOT ladder alone therefore does not ensure
that the complete Trotter-plus-QEC experiment falls below the
reference line $P_L=p_{\mathrm{phase}}$.

\begin{figure}[t]
    \centering
    \includegraphics[
        width=\columnwidth
    ]{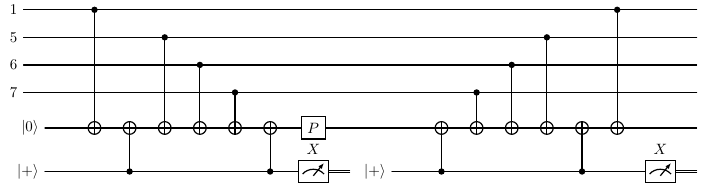}
    \caption{
        Revised two-flag circuit used in the diagnostic protected
        limit. Relative to the original placement in
        Fig.~\ref{fig:833_flagged_trotter_circuit}, the flag windows
        are shifted so that the outermost parity-network CNOTs are
        not enclosed by the flags. The phase gate and the CNOTs
        belonging to the flag gadgets are idealized only in this
        diagnostic experiment.
    }
    \label{fig:833_revised_diagnostic_flag_circuit}
\end{figure}

Finally, to understand a setting where both logical sectors are suppressed
and exhibit a pseudo-threshold, we use the single-fault propagation analysis to construct a
diagnostic protected limit. The phase gate and the CNOTs belonging to
the two flag gadgets are treated as ideal, while the remaining
Trotter CNOTs are followed by the restricted $Z$-biased channel in
Eq.~\eqref{eq:833_z_biased_two_qubit_channel}. The flag placement is
chosen so that, within this restricted single-fault model, the
trivial-syndrome flag record does not identify both the data identity
and a nontrivial logical operator. The revised placement, shown in
Fig.~\ref{fig:833_revised_diagnostic_flag_circuit}, excludes the
outermost parity-network CNOTs from the two flag windows. This circuit
is used only for the diagnostic protected limit and does not replace
the original flag circuit used in the preceding fault-enumeration and
postselection experiments.

The recovery is branch-conditioned. If neither flag is triggered, we
apply one noisy Chao--Reichardt error-correction round. If exactly one
flag is triggered, we perform one noisy full unflagged
syndrome-extraction pass and decode the resulting five-bit syndrome
using the flag-conditioned Trotter lookup table. Double-flag events
are discarded. Every accepted shot is then followed by the final
ideal cleanup. The remaining noisy Trotter CNOTs and all noisy
locations in the branch-conditioned recovery use the same swept
error parameter $p$; the Trotter CNOT channel is restricted to the
$Z$-biased model, whereas the recovery circuits retain their
depolarizing circuit-level noise model.

\begin{figure}[t]
    \centering
    \includegraphics[
        width=\columnwidth
    ]{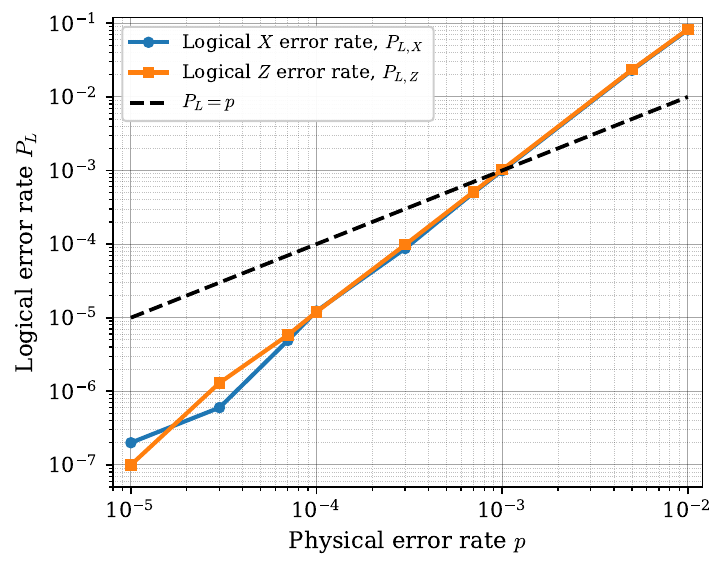}
    \caption{
        Diagnostic protected limit with ideal phase and flag-gadget
        CNOTs and $Z$-biased noise on the remaining Trotter CNOTs.
        Recovery is branch conditioned: the no-flag branch invokes
        one noisy Chao--Reichardt round, the single-flag branch invokes
        one noisy full unflagged syndrome-extraction pass followed by
        flag-conditioned decoding, and double-flag events are
        discarded. Every accepted shot receives final ideal cleanup.
        Both logical sectors exhibit a diagnostic pseudo-threshold
        relative to $P_L=p$. Each point is estimated from $N=10^{7}$
        attempted Monte Carlo shots; only point estimates are shown.
    }
    \label{fig:833_ideal_flag_phase_limit}
\end{figure}

Under these assumptions, both logical sectors exhibit a diagnostic
pseudo-threshold relative to the reference line $P_L=p$, as shown in
Fig.~\ref{fig:833_ideal_flag_phase_limit}. The low-noise dependence
is consistent with the removal of the malignant first-order Trotter
faults present in the preceding implementations. The resulting
curves cross the reference line on the same
$p\sim10^{-3}$ scale as the circuit-level memory benchmark in
Fig.~\ref{fig:833_rui_memory}. This agreement is consistent with the
interpretation that the dominant remaining contribution is associated
with the noisy recovery circuit rather than with the protected
Trotter block.

This experiment should not be interpreted as a hardware-ready
fault-tolerant implementation. Rather, it identifies the smallest
set of idealizations tested in this work under which the malignant
first-order Trotter mechanisms are removed and both logical sectors
recover threshold-like behavior. Relaxing the ideal phase and
flag-gadget assumptions while retaining this suppression is the
central circuit-design problem left open.

At the smallest physical error probabilities, only a small number of
logical failures are observed even with $N=10^{7}$ shots. The
corresponding exact binomial intervals are therefore relatively broad
on a logarithmic scale. These points support suppression below the
reference line, but they are not used to extract a precise
low-noise prefactor or scaling exponent.

\section{Implementation implications and small-code overhead}
\label{sec:833_implementation}

The asymmetric-noise experiments suggest that the CNOT backbone of
the encoded Trotter circuit must operate at an error rate
substantially below that of an unprotected two-qubit implementation.
One possible route is to implement the required CNOTs through
high-fidelity Bell pairs and gate teleportation~\cite{gottesman1999quantum,feng2025chip}. Alternatively, the physical coordinates of the outer $\llbr 8,3,3\rrbr$ block could be
protected by a small inner code.

As a representative small-code construction, concatenating each of
the eight outer-code coordinates with a distance-three rotated
surface-code patch~\cite{fowler2012surface,tomita2014low} gives
\begin{equation}
    N_{\mathrm{data}}
    =
    8\times9
    =
    72
    \label{eq:833_surface_concatenation_data_count}
\end{equation}
surface-code data qubits. This number counts only the nine data
qubits in each $\llbr 9,1,3\rrbr$ inner patch. It does not include
the measurement ancillas required by the inner surface-code
syndrome-extraction circuits, the ancillas used for outer-code error
correction, the Trotter parity ancilla, or the flag qubits.
Consequently, Eq.~\eqref{eq:833_surface_concatenation_data_count}
is a code-block data-qubit count rather than a complete hardware
footprint.

In particular, the value $72$ counts only storage of the eight
outer-code data coordinates. If the common Trotter parity target is
also represented by a distance-three inner-code patch, a ninth patch
is required, increasing the active inner-code data-qubit count to
$9\times9=81$ before syndrome ancillas are included. Alternatively,
if the parity target remains bare, a separate fault-tolerant
encoded-to-bare CNOT interface must be specified. Neither option is
included in the $72$-qubit count.

The purpose of this concatenated picture is not to provide a
full architecture-level resource comparison. Instead, it illustrates
one mechanism by which the effective CNOT error rate assumed in
Sec.~\ref{sec:833_asymmetric_noise} could be approached. The actual
logical-CNOT error rate would depend on the syndrome-extraction
schedule, decoder, number of correction rounds, connectivity, and
implementation of the inter-patch CNOT.

Moreover, suppressing the CNOT error rate alone does not remove the
first-order contribution from the central phase rotation. The
asymmetric benchmark in
Fig.~\ref{fig:833_asymmetric_direct_trotter} explicitly demonstrates
this limitation: once the CNOT contribution is reduced, the
logical-$Z$ failure rate becomes controlled by the phase-gate error.
The concatenated construction can therefore motivate the
$p_{\mathrm{CNOT}}=10^{-5}$ regime, but it does not by itself provide
simultaneous fault tolerance for the complete encoded rotation.

Concatenation can suppress the CNOT-induced error floor and may
therefore create a parameter regime in which a pseudo-threshold
becomes possible. It is not sufficient by itself: if the central
phase rotation remains unprotected, the logical-$Z$ failure
probability retains the first-order contribution identified in
Eq.~\eqref{eq:833_asymmetric_phase_scaling}. A pseudo-threshold in
both sectors would consequently require concatenation to be combined
with a complementary mechanism that suppresses or detects dangerous
phase-gate faults. The logical failure probability need not vanish;
the relevant criterion is whether the complete encoded
implementation falls below a consistently defined physical or
unencoded reference.

Since the present study concerns a single small non-CSS code block,
we do not attempt a direct space--time comparison with large-scale
architectures. Such a comparison would require matching logical
accuracy, state-preparation cost, syndrome-extraction depth, decoder
latency, connectivity, and postselection overhead rather than only
the number of data qubits. Our quantitative comparisons are therefore
restricted to the unencoded rotation, the direct encoded circuit,
the Chao--Reichardt memory benchmark, CliNR, and flag
postselection under common simulation conventions.

\section{Discussion and open problems}
\label{sec:discussion}

The circuit-level experiments above identify several limitations that
are specific to implementing a logical Pauli rotation on a small
high-rate non-CSS code. These limitations arise from the terminal
recovery convention, the preparation of the encoded input states, the
dynamic structure of the Chao--Reichardt recovery procedure, and the
propagation of faults through the common parity ancilla. We discuss
these issues below and distinguish the conclusions established by the
simulations from the assumptions that remain to be removed in a
hardware-realistic implementation.

\paragraph{Terminal cleanup and the definition of logical failure.}

All circuit-level logical-gate experiments use a final ideal cleanup
before the propagated logical observables are measured. Its purpose is
to distinguish a correctable weight-one residual error from a
nontrivial logical error at the end of the simulated circuit. Without
this terminal convention, a correctable error left by the final noisy
error-correction round can affect the reported logical measurement
even though it has not yet produced an encoded failure.

For CSS codes, terminal decoding can often be incorporated naturally
into a final product-basis measurement. The separation between
\(X\)- and \(Z\)-type checks permits the corresponding syndrome and
logical parity information to be inferred from measurements in a
common basis. The same simplification does not extend directly to the
non-CSS \(\llbr 8,3,3\rrbr\) code. Its stabilizers and logical
operators contain overlapping \(X\), \(Y\), and \(Z\) components, so
a single product-basis measurement does not generally provide all of
the mixed-Pauli information required for terminal decoding.

A related boundary dependence was studied for the non-CSS
\(\llbr 8,1,3\rrbr\) code in
Ref.~\cite{maheshwari2024fault}. There, a noise-free final
detection-and-correction projection removed residual correctable
errors that otherwise contributed to the finite-circuit output
statistics. This supports the use of an ideal terminal projection as
a diagnostic simulation convention, but it does not provide a
hardware-realistic implementation of that projection.

A practical replacement would require additional noisy
syndrome-extraction rounds followed by a spacetime decoder. For the
present protocol, this is complicated by the fact that the recovery
circuit is dynamic: earlier measurement outcomes determine which
extraction circuits are subsequently executed. A suitable terminal
decoder must therefore infer not only data and measurement faults, but
also whether the observed classical record caused the circuit to
follow the correct recovery branch. Developing such a decoder remains
an open problem.

For a fixed syndrome-extraction schedule, the circuit can in
principle be compiled into a static detector error model (DEM)~\cite{gidney2021stim,derks2407designing}. The complete spacetime syndrome record could then be decoded jointly~\cite{dennis2002topological}, rather than reducing each extraction round independently to a lookup table entry. Such a formulation could also incorporate noisy terminal measurements into the decoding problem and thereby provide a route toward replacing the final ideal-cleanup convention.

The dynamic Chao--Reichardt procedure does not directly admit one
conventional static DEM because different measurement records select
different circuits, detector sets, and causal relations. Possible
extensions include compiling each recovery branch into a separate
DEM, embedding all branches into a common padded measurement
schedule, or conditioning the decoder explicitly on the complete
classical branch record. Whether any of these representations can
remove the need for terminal ideal cleanup remains open.

\paragraph{Idealized state preparation.}

The encoded state-preparation convention is also idealized. The five
stabilizer generators are measured projectively, and the
corresponding weight-one recovery from
\(\mathcal{D}_{\mathrm{LUT}}\) is applied whenever the measured
syndrome belongs to the weight-one table. A trivial syndrome requires
no recovery, whereas a syndrome outside the table causes the
preparation attempt to be rejected and restarted. The required
logical eigenvalues are then imposed through ideal projective
measurements of the logical Pauli operators.

This procedure prepares a well-defined input ensemble and isolates
the performance of the subsequent noisy circuit. It should not,
however, be interpreted as a fault-tolerant state-preparation
protocol. A hardware implementation could replace the ideal
projection by repeated noisy extraction with verification or
postselection, but measurement faults could cause an incorrect state
to be accepted, and the rejection probability could become
substantial.

General graphical and unitary encoders are known for stabilizer codes,
including constructions associated with the Gottesman eight-qubit
code~\cite{cafaro2014scheme,cafaro2022geometric}. Such ideal encoding
circuits do not automatically provide fault-tolerant preparation,
because a single encoder fault can propagate to multiple data
qubits. Verified encoding, automated preparation synthesis, and
fault-tolerant code conversion provide possible starting
points~\cite{peham2025automated,anderson2014fault}, but a
low-overhead preparation procedure specialized to the
\(\llbr 8,3,3\rrbr\) logical ensembles remains to be developed.

\paragraph{Dynamic recovery and circuit latency.}

The Chao--Reichardt procedure is intrinsically dynamic. A stabilizer
is first measured using its flagged extraction circuit, after which
the flag and syndrome outcomes determine whether the protocol
continues to the next flagged generator or invokes a complete set of
unflagged syndrome-extraction circuits. Consequently, the circuit
depth, number of entangling gates, and number of measurements are
random variables that depend on the observed measurement record.

This dynamic structure has both physical and computational
consequences. On hardware, mid-circuit measurements must be processed
before the next recovery branch can be selected. Measurement,
classical decoding, feedforward, ancilla reset, and conditional
circuit loading can therefore introduce latency beyond the nominal
quantum-gate depth. The impact is particularly important when
measurement and reset are substantially slower than single- or
two-qubit gates.

The same structure makes Monte Carlo simulation more expensive.
Different shots can execute different circuits, so the complete
experiment cannot always be represented by one fixed stabilizer
circuit and sampled in a single large batch. Shot-dependent circuit
construction and classical branching reduce the advantage normally
provided by highly vectorized stabilizer simulation. The runtime
observed in the present simulations is therefore partly a consequence
of the protocol's dynamic control flow rather than only the number of
physical qubits.

Dynamic execution is not itself a correctness problem, but it is an
important implementation cost. A practical version of the protocol
would benefit from replacing physical corrections by Pauli-frame
updates, compiling as much of the branch structure as possible into a
fixed schedule, and using a decoder that processes the entire
measurement history without repeatedly changing the quantum circuit.
Whether the Chao--Reichardt procedure can be reformulated in this way
for the non-CSS \(\llbr 8,3,3\rrbr\) code remains open.

\paragraph{First-order limitations of the encoded Trotter circuit.}

The main limitation of the direct encoded-Trotter implementation is
not the distance of the underlying code in isolation, but the
circuit-level propagation of a fault on the auxiliary parity qubit.
A \(Z\) component produced after the central phase gate can traverse
the parity-uncomputation ladder and become the complete data operator
\(\overline{Z_2}=Z_1Z_5Z_6Z_7\). Because this operator has trivial
stabilizer syndrome, neither the subsequent Chao--Reichardt recovery
nor the final weight-one cleanup can distinguish it from the
identity. So the direct Trotter gadget has circuit-level
distance $1$ in this logical sector.

The asymmetric-noise experiment separates this phase-gate mechanism
from faults in the CNOT backbone. When
\(p_{\mathrm{CNOT}}=10^{-5}\) is fixed and
\(p_{\mathrm{phase}}\) is varied, the high-noise logical-\(Z\)
behavior is approximately
\(P_{L,Z}\simeq2p_{\mathrm{phase}}/3\). This is consistent with the
two phase-containing components, \(Y\) and \(Z\), of the
single-qubit depolarizing channel following the phase gate. At small
\(p_{\mathrm{phase}}\), this contribution no longer dominates and
the logical failure rates approach a floor set by the fixed noisy
CNOT locations.

Thus, improving the Clifford backbone does not remove the first-order
contribution of the analog rotation. It changes which physical
operation limits the encoded logical error rate. Conversely,
protecting only the phase rotation would leave the correlated faults
introduced by the CNOT ladder. Simultaneous suppression requires the
two components to be treated jointly.

Adding one noisy Chao--Reichardt round makes the asymmetric result
worse because the recovery circuit contributes additional noisy
entangling locations. This observation does not imply that error
correction is intrinsically harmful. Rather, it shows that reducing
the error probability of the Trotter CNOTs alone is insufficient when
the subsequent syndrome-extraction circuit continues to operate at a
larger physical error scale.

\paragraph{Relation to pieceable fault tolerance.}

Pieceable fault tolerance suggests a natural response to long
propagation paths: divide a nontransversal logical interaction into
pieces and perform intermediate error correction before a single fault
can spread through the complete gate~\cite{yoder2016universal,
yoder2018practical}. Its applicability depends on retaining a
measurable stabilizer description at each intermediate boundary. For
example, in suitable controlled-$Z$ constructions, relevant pure-
$Z$ checks commute with the intervening CZ gates and remain constant
Pauli stabilizers, enabling intermediate syndrome extraction.

As already observed by Yoder for an analogous logical Pauli-rotation
circuit without functional ancillas~\cite{yoder2018practical}, the single-block Trotter circuit does not generally have this
property. After a partial evolution $U_t$, the state occupies the
transformed codespace $U_t\mathcal C$, stabilized by
$U_t\mathcal S U_t^\dagger$, rather than the original codespace
$\mathcal C$. For a general-angle Pauli rotation, stabilizers that
anticommute with the rotation are conjugated into non-Pauli linear
combinations. Measuring the original checks would therefore disturb
the intended evolution, while standard Pauli syndrome extraction
cannot directly measure the transformed checks. At Clifford rotation
angles the transformed checks remain Pauli, so a tailored pieceable
construction may still be possible, but it must also protect the
central rotation and the final circuit segment. Pieceable fault
tolerance is therefore a relevant design direction, not an immediate
drop-in remedy for the Trotter gadget studied here.

\paragraph{Limits of flag-conditioned recovery.}

The exhaustive single-fault analysis shows that the complete record
\((f_{\mathrm L},f_{\mathrm R},\boldsymbol{s})\) does not always
identify a unique logical coset of the residual data error. Distinct
faults can produce identical flag outcomes and identical stabilizer
syndromes while leaving data errors that differ by a nontrivial
logical operator. Such a collision cannot be resolved by merely
adding more entries to a lookup table.

The trivial-syndrome sector gives the clearest example. Under the
original flag placement, the same flagged record can be compatible
with either the identity on the data block or
\(\overline{Z_2}\). Applying \(\overline{Z_2}\) as recovery corrects
one case but converts the identity case into a logical error; applying
the identity has the opposite failure mode. A probabilistic recovery
can change the coefficient of the resulting failure probability but
cannot remove its first-order contribution.

The revised diagnostic flag placement excludes the outermost CNOT
locations responsible for this particular identity--logical
collision under the restricted \(Z\)-biased single-fault model.
Errors at the excluded locations either leave the data unchanged or
produce a correctable weight-one error that can be handled by final
cleanup. This modification removes one specific ambiguity, but it
does not establish fault tolerance under a general noisy flag model.
When the flag couplings are noisy, false alarms, missed detections,
and additional hook errors remain possible.

The \(Z\)-type flags reported here were selected after examining
alternative placements and observables. An \(X\)-sensitive
construction can detect some faults transformed by the phase gate,
but requires additional entangling operations. In the tested
circuits, the information gained from those flags did not compensate
for the additional noisy locations. This illustrates a general
constraint of flag design: the flag interactions must preserve the
intended logical circuit while detecting dangerous propagation, and
that equivalence requirement sharply limits where additional
couplings can be inserted.

\paragraph{Interpretation of the diagnostic protected limit.}

The experiment with ideal phase and flag-gadget CNOTs identifies a
restricted regime in which both logical sectors exhibit a diagnostic
pseudo-threshold relative to $P_L=p$. Within that model, the remaining Trotter
CNOTs are subject to \(Z\)-biased noise, while the flag geometry and
recovery remove the malignant first-order Trotter faults considered
in the single-fault analysis. The resulting curves are close to the
circuit-level memory scale.

This agreement provides a useful diagnostic conclusion: once the
identified first-order Trotter mechanisms are removed, the dominant
logical failure probability is no longer set by the encoded rotation
but by the noisy Chao--Reichardt memory-protection circuit. In this
restricted model, the encoded Trotter layer therefore adds primarily
higher-order contributions to the memory-level failure rate.

The result should not be interpreted as a hardware-ready protocol.
The ideal phase gate removes precisely the physical location that
dominates the logical-\(Z\) channel of the direct implementation, and
ideal flag couplings eliminate faults introduced by the detector
itself. The experiment instead identifies the smallest set of
idealizations tested in this work under which the known first-order
malignant mechanisms are absent.

A realistic extension could use high-fidelity teleported CNOTs to
reduce faults in the parity network and flag gadgets. This may make
hook errors from those locations sufficiently rare, but it does not
by itself protect the central arbitrary-angle rotation. Removing the
ideal phase assumption without reintroducing a first-order logical
operator remains the principal unresolved component of the
construction. An encouraging fact is that in some trapped-ion systems,
such $Z$-rotations are realized via classical phase tracking and thus
have a zero error rate~\cite{Quantinuum-H2}. In such cases, the only remaining ideality is
the assumption of perfect CNOTs in the flag gadgets. If we perform 
those as well using high-fidelity Bell pairs, then we could potentially
turn this diagnostic setting into a practical protocol.

The results also suggest that no single mitigation mechanism studied
here addresses all dominant fault classes. A more promising direction
is a hybrid construction in which high-fidelity teleported or
inner-code-protected CNOTs suppress parity-network faults, while a
complementary phase-sensitive detection or verification primitive
targets faults introduced by the central rotation. A
branch-conditioned spacetime decoder could then combine the flag
record, stabilizer outcomes, and terminal noisy measurements instead
of assigning corrections through independent lookup tables. Such a
hybrid strategy may retain a nonzero logical error rate while still
providing a genuine pseudo-threshold in both logical sectors.

\paragraph{Axis dependence of CliNR and flag postselection.}

CliNR and direct flag postselection use different rejection
mechanisms. CliNR verifies an auxiliary resource state before it
interacts with the input data, whereas flag postselection rejects a
shot after the encoded computation has already begun. Nevertheless,
the accepted error ensembles exhibit a similar dependence on the
rotation axis.

Resource checks or flag observables aligned with a \(Z\)-type
rotation strongly suppress many transverse faults and can reduce the
logical-\(X\) failure rate by several orders of magnitude, while
logical-\(Z\) components aligned with the rotation axis survive. For
an \(X\)-type rotation, the roles of the two logical sectors are
approximately reversed. This similarity suggests that the observed
asymmetry is not solely a defect of either postselection rule. It is
closely connected to the conjugation structure of the implemented
Pauli rotation.

CliNR has the advantage that an unsuccessful resource state can be
rejected offline before consuming the input state. Its limitation in
the present setting is that the complete \(\pi/4\) Trotter block must
be Clifford. Applying CliNR separately to the two CNOT ladders would
permit a general intermediate rotation, but the split construction
tested here introduced additional teleportation boundaries and
performed worse than full-block CliNR.

Flag postselection can surround a genuinely non-Clifford rotation,
but its rejection occurs online. If a single block has acceptance
probability \(P_{\mathrm{acc}}\), repeating the procedure over \(L\)
blocks gives an approximate survival probability
\(P_{\mathrm{acc}}^L\). Even a favorable single-block acceptance rate
can therefore lead to substantial sampling overhead in a long
simulation. Moreover, rejecting every triggered flag discards benign
and correctable faults together with damaging correlated errors.

The present results consequently do not identify either method as a
complete fault-tolerant replacement for direct execution. They show
instead that both resource verification and circuit-level flagging can
strongly suppress one logical sector and provide information about
the dominant propagation mechanisms.

\paragraph{Small-code implementation scope.}

One possible method for reducing the effective CNOT error rate is to
concatenate each physical coordinate of the
\(\llbr 8,3,3\rrbr\) block with a distance-three rotated
surface-code patch. Counting the nine data qubits in each inner
\(\llbr 9,1,3\rrbr\) patch gives \(8\times9=72\) inner-code data
qubits. This is not the complete physical-qubit footprint: it excludes
the syndrome ancillas of the inner patches, the outer-code extraction
ancillas, the Trotter parity ancilla, and all flag qubits.

If the common parity target is also encoded into an inner patch, the
active inner-code data count becomes $81$ rather than $72$. If it is
left bare, a fault-tolerant encoded-to-bare CNOT interface is instead
required. Thus, the $72$-qubit value is only a static outer-data-block
count and not a complete logical-gate implementation cost.

Such concatenation provides a possible route toward the strongly
suppressed CNOT scale used in the asymmetric benchmark, but it does
not automatically produce a pseudo-threshold for the complete
logical rotation. The phase-gate contribution remains first order
unless the rotation itself is protected or its dangerous faults are
detected before parity uncomputation.

For this reason, we do not present the \(72\)-data-qubit count as a
direct resource advantage over an existing architecture. A meaningful
comparison would require matching logical accuracy and including
syndrome-extraction depth, decoder latency, measurement ancillas,
state preparation, connectivity, and the cost of the logical CNOT
implementation. The quantitative comparisons in this work are
therefore restricted to the unencoded circuit, the direct encoded
implementation, the Chao--Reichardt memory benchmark, CliNR, and flag
postselection under consistent simulation conventions.

\section{Conclusion and outlook}
\label{sec:conclusion}

We have presented a circuit-level study of logical Trotterization
with the non-CSS \(\llbr 8,3,3\rrbr\) code, progressing from
code-capacity and phenomenological benchmarks to noisy memory
protection and encoded logical rotations. Under the stated noise and
terminal-cleanup conventions, the Chao--Reichardt memory experiment
exhibits a pseudo-threshold near \(1.5\times10^{-3}\), demonstrating
that this small, high-rate code can provide threshold-like
suppression when used as a quantum memory.

Logical Trotter execution introduces a qualitatively different
failure mechanism. A single phase-containing fault on the common
parity ancilla can propagate through the uncomputation ladder into a
full logical-\(Z\) operator with trivial stabilizer syndrome. The
direct gadget therefore has circuit-level distance $1$ in this
sector even though the underlying code has distance $3$. CliNR and
flag postselection suppress important subsets of the propagated
faults, but their performance remains strongly axis dependent, with
the logical sector aligned with the rotation remaining the more
difficult one.

The asymmetric-gate-noise benchmarks separate the principal
first-order contributions. At fixed
\(p_{\mathrm{CNOT}}=10^{-5}\), the low-noise behavior is limited by a
CNOT-induced floor, whereas the high-noise logical-\(Z\) failure rate
scales approximately as \(2p_{\mathrm{phase}}/3\). Adding a noisy
Chao--Reichardt recovery round increases the failure rate through its
additional circuit locations. By contrast, in the diagnostic
protected limit---where the central phase operation and flag
couplings are ideal and the remaining Trotter CNOTs are subject to
\(Z\)-biased noise---both logical sectors exhibit diagnostic
pseudo-thresholds relative to \(P_L=p\). The resulting curves closely
track the memory benchmark, indicating that once the identified
first-order Trotter faults are removed, memory protection becomes the
dominant limitation. This diagnostic result identifies the protection
needed for two-sector suppression but does not constitute a
hardware-ready protocol.

Taken together, these results show that preserving the logical
Trotter pattern under encoding does not by itself preserve
fault-tolerant distance. A complete construction requires the parity
geometry, analog rotation, flag placement, syndrome-extraction
schedule, and classical recovery rule to be designed jointly so that
every relevant single fault either remains correctable or produces a
distinguishable measurement record. It must also replace the ideal
phase operations, ideal flag couplings, terminal ideal cleanup, and
online rejection used in the diagnostic limit with noisy,
fault-tolerant procedures.

The \(\llbr 8,3,3\rrbr\) code serves as a useful circuit-level
testbed because its small size permits explicit fault enumeration and
direct identification of syndrome collisions. In parallel, we are
investigating the same logical-Trotter construction on finite-rate
QLDPC code families
~\cite{panteleev2022asymptotically,raveendran2022finite}, where
increasing distance, redundant syndrome information, and spacetime
decoding may alter the balance between local-rotation faults and
parity-network faults. The mechanisms isolated here provide concrete
diagnostic tests for those larger-code simulations: whether ancilla
phase faults remain first order, whether their propagated errors
retain distinguishable syndromes, and how both effects change with
code distance. Hybrid CNOT protection, phase-sensitive verification,
static or branch-conditioned DEM decoding, and scaling studies on
QLDPC code families provide concrete directions for extending the
present small-code analysis.

\begin{acknowledgments}
The authors thank David Hayes, Nicolas Delfosse, James Brown, and Rui Chao
for insightful discussions.
Chen also thanks Mingyuan Wang for providing a fast Julia implementation of 
the flagged syndrome extraction procedure under circuit-level noise.
This work was supported in part by the University of Arizona Office of Research and Partnerships through the Bridge Funding Investment Program and by the U.S. National Science
Foundation through the NQVL program under Award No.~2547483. 
Any opinions, findings,
conclusions, or recommendations expressed in this material are those
of the authors and do not necessarily reflect the views of the
National Science Foundation.

The authors used OpenAI's ChatGPT (GPT-5.6) to assist with code
generation and debugging, analytical cross-checks, manuscript
organization, and language and grammar editing. All AI-assisted code,
analyses, and text were independently reviewed, tested, and validated
by the authors. The authors take full responsibility for the
scientific content, interpretation of the results, and final
manuscript.
\end{acknowledgments}

\appendix

\section{Weight-one lookup table for the
         $\llbr 8,3,3\rrbr$ code
}
\label{app:833_weight1_lut}

Table~\ref{tab:833_weight1_lut} lists the syndromes of all
single-qubit Pauli errors for the $\llbr 8,3,3\rrbr$ code. Syndrome
bits are ordered according to the extraction generators
$(g_1,g_2,g_3,g_4,g_5)$ defined in
Eq.~\eqref{eq:833_extraction_generators}. This is the same syndrome
convention used in the code-capacity, phenomenological-noise, and
circuit-level simulations.

Because the code is non-degenerate, all 24 single-qubit Pauli errors
have distinct nonzero syndromes. Since Pauli operators are
self-inverse up to a physically irrelevant global phase, the decoder
applies the corresponding Pauli operator as the recovery associated
with each syndrome.

\begin{table}[ht]
    \centering
    \caption{
        Syndromes of single-qubit Pauli errors for the
        $\llbr 8,3,3\rrbr$ code. Syndrome bits are ordered as
        $(g_1,g_2,g_3,g_4,g_5)$, using the extraction generators
        defined in Eq.~\eqref{eq:833_extraction_generators}. For each
        syndrome, the decoder applies the corresponding Pauli
        operator as the recovery.
    }
    \label{tab:833_weight1_lut}
    \begin{tabular}{cccc}
        \toprule
        Qubit
        &
        $\boldsymbol{s}(X_i)$
        &
        $\boldsymbol{s}(Y_i)$
        &
        $\boldsymbol{s}(Z_i)$
        \\
        \midrule
        $1$ & $01000$ & $11000$ & $10000$ \\
        $2$ & $01010$ & $11011$ & $10001$ \\
        $3$ & $10100$ & $00110$ & $10010$ \\
        $4$ & $10101$ & $11001$ & $01100$ \\
        $5$ & $01001$ & $00111$ & $01110$ \\
        $6$ & $10110$ & $00101$ & $10011$ \\
        $7$ & $10111$ & $11010$ & $01101$ \\
        $8$ & $01011$ & $00100$ & $01111$ \\
        \bottomrule
    \end{tabular}
\end{table}

\section{Flag-conditioned lookup tables}
\label{app:833_flag_luts}

When the flag is triggered during the extraction of $g_i$, the
complete syndrome is supplemented by the identity $i$ of the
triggered extraction circuit. The decoder therefore uses the pair
$(i,\boldsymbol{s})$ to select the corresponding recovery from
Table~\ref{tab:833_flag_luts}.

The controlled-Pauli interactions in
Fig.~\ref{fig:833_flagged_circuits} are ordered so that the correlated
errors associated with a triggered flag have distinct complete
syndromes within each individual extraction circuit. Errors belonging
to different extraction circuits need not have globally distinct
syndromes because the circuit index $i$ is retained as part of the
decoder input.

Since Pauli operators are self-inverse up to a physically irrelevant
global phase, each entry in Table~\ref{tab:833_flag_luts} lists the
inferred correlated Pauli error itself as the corresponding recovery
operator.

\begin{table*}
    \centering
    \caption{
        Flag-conditioned lookup tables for the five flagged
        syndrome-extraction circuits. Syndrome bits are ordered as
        $(g_1,g_2,g_3,g_4,g_5)$, using the extraction generators
        defined in Eq.~\eqref{eq:833_extraction_generators}. The
        identity of the triggered extraction circuit is retained as
        part of the decoder input.
    }
    \label{tab:833_flag_luts}

    \footnotesize
    \setlength{\tabcolsep}{4pt}


    \begin{minipage}[t]{0.31\textwidth}
        \vspace{0pt}
        \centering
        \textbf{(a) Flag-conditioned LUT for $g_1$}

        \smallskip

        \begin{tabular}{cc}
            \toprule
            Correction & Syndrome \\
            \midrule

            $X_4Z_7$ & $11000$ \\
            $Y_4Z_7$ & $10100$ \\
            $Z_4Z_7$ & $00001$ \\
            \midrule

            $Z_4X_6Z_7$ & $10111$ \\
            $Z_4Y_6Z_7$ & $00100$ \\
            $Z_4Z_6Z_7$ & $10010$ \\
            \midrule

            $X_3Z_4Y_6Z_7$ & $10000$ \\
            $Y_3Z_4Y_6Z_7$ & $00010$ \\
            $Z_3Z_4Y_6Z_7$ & $10110$ \\
            \midrule

            $X_2Y_3Z_4Y_6Z_7$ & $01000$ \\
            $Y_2Y_3Z_4Y_6Z_7$ & $11001$ \\
            $Z_2Y_3Z_4Y_6Z_7$ & $10011$ \\

            \bottomrule
        \end{tabular}
    \end{minipage}
    \hfill
    \begin{minipage}[t]{0.31\textwidth}
        \vspace{0pt}
        \centering
        \textbf{(b) Flag-conditioned LUT for $g_2$}

        \smallskip

        \begin{tabular}{cc}
            \toprule
            Correction & Syndrome \\
            \midrule

            $X_7Y_8$ & $10011$ \\
            $Y_7Y_8$ & $11110$ \\
            $Z_7Y_8$ & $01001$ \\
            \midrule

            $X_4X_7Y_8$ & $00110$ \\
            $Y_4X_7Y_8$ & $01010$ \\
            $Z_4X_7Y_8$ & $11111$ \\
            \midrule

            $X_4X_5X_7Y_8$ & $01111$ \\
            $X_4Y_5X_7Y_8$ & $00001$ \\
            $X_4Z_5X_7Y_8$ & $01000$ \\
            \midrule

            $X_2X_4Y_5X_7Y_8$ & $01011$ \\
            $Y_2X_4Y_5X_7Y_8$ & $11010$ \\
            $Z_2X_4Y_5X_7Y_8$ & $10000$ \\

            \bottomrule
        \end{tabular}
    \end{minipage}
    \hfill
    \begin{minipage}[t]{0.31\textwidth}
        \vspace{0pt}
        \centering
        \textbf{(c) Flag-conditioned LUT for $g_3$}

        \smallskip

        \begin{tabular}{cc}
            \toprule
            Correction & Syndrome \\
            \midrule

            $X_6Y_7$ & $01100$ \\
            $Y_6Y_7$ & $11111$ \\
            $Z_6Y_7$ & $01001$ \\
            \midrule

            $Z_6Y_7X_8$ & $00010$ \\
            $Z_6Y_7Y_8$ & $01101$ \\
            $Z_6Y_7Z_8$ & $00110$ \\
            \midrule

            $X_4Z_6Y_7X_8$ & $10111$ \\
            $Y_4Z_6Y_7X_8$ & $11011$ \\
            $Z_4Z_6Y_7X_8$ & $01110$ \\
            \midrule

            $Y_4X_5Z_6Y_7X_8$ & $10010$ \\
            $Y_4Y_5Z_6Y_7X_8$ & $11100$ \\
            $Y_4Z_5Z_6Y_7X_8$ & $10101$ \\

            \bottomrule
        \end{tabular}
    \end{minipage}

    \par\medskip


    \makebox[\textwidth][c]{%
        \begin{minipage}[t]{0.31\textwidth}
            \vspace{0pt}
            \centering
            \textbf{(d) Flag-conditioned LUT for $g_4$}

            \smallskip

            \begin{tabular}{cc}
                \toprule
                Correction & Syndrome \\
                \midrule

                $X_3X_5$ & $11101$ \\
                $Y_3X_5$ & $01111$ \\
                $Z_3X_5$ & $11011$ \\
                \midrule

                $X_3X_5X_8$ & $10110$ \\
                $X_3X_5Y_8$ & $11001$ \\
                $X_3X_5Z_8$ & $10010$ \\
                \midrule

                $X_3X_5X_7Y_8$ & $01110$ \\
                $X_3X_5Y_7Y_8$ & $00011$ \\
                $X_3X_5Z_7Y_8$ & $10100$ \\
                \midrule

                $X_3X_5X_6Z_7Y_8$ & $00010$ \\
                $X_3X_5Y_6Z_7Y_8$ & $10001$ \\
                $X_3X_5Z_6Z_7Y_8$ & $00111$ \\

                \bottomrule
            \end{tabular}
        \end{minipage}
        \hspace{0.04\textwidth}
        \begin{minipage}[t]{0.31\textwidth}
            \vspace{0pt}
            \centering
            \textbf{(e) Flag-conditioned LUT for $g_5$}

            \smallskip

            \begin{tabular}{cc}
                \toprule
                Correction & Syndrome \\
                \midrule

                $Y_7X_8$ & $10001$ \\
                $Y_7Y_8$ & $11110$ \\
                $Y_7Z_8$ & $10101$ \\
                \midrule

                $X_6Y_7Y_8$ & $01000$ \\
                $Y_6Y_7Y_8$ & $11011$ \\
                $Z_6Y_7Y_8$ & $01101$ \\
                \midrule

                $X_5X_6Y_7Y_8$ & $00001$ \\
                $Y_5X_6Y_7Y_8$ & $01111$ \\
                $Z_5X_6Y_7Y_8$ & $00110$ \\
                \midrule

                $X_4Z_5X_6Y_7Y_8$ & $10011$ \\
                $Y_4Z_5X_6Y_7Y_8$ & $11111$ \\
                $Z_4Z_5X_6Y_7Y_8$ & $01010$ \\

                \bottomrule
            \end{tabular}
        \end{minipage}%
    }

\end{table*}


\bibliography{references}

\end{document}